\documentclass[11pt]{article}
\usepackage[body={16cm,23cm}]{geometry}

\usepackage[hang,small,center]{caption}
\usepackage{amsmath,amssymb}
\usepackage{amsfonts}
\usepackage{mathrsfs}
\usepackage{epic}
\usepackage{eepic}
\usepackage{graphicx}
\usepackage{cite}
\newcommand{\ds}{\displaystyle}
\font\cyr=wncyr10 
\newcommand{\Lb}{\operatorname{\mbox{\cyr L}}}
\renewcommand{\author}[1]{\large\rm #1\\ \bigskip}
\newcommand{\address}[1]{{\normalsize\it #1\\}\bigskip}
\renewcommand{\title}[1]{\bigskip\bigskip\Large\bf #1\bigskip\bigskip\\}

\newcommand{\Bigpsi}[3]{\phantom{\Psi}_2 \kern -.05em
\Psi_2\left(\genfrac{}{}{0pt}{}{#1}{#2}\biggl|#3\right)}

\newcommand{\be}{\begin{equation}}
\newcommand{\ee}{\end{equation}}

\newcommand{\tvec}[2]{\genfrac{(}{)}{0pt}{}{#1}{#2}}
\newcommand{\x}{{\boldsymbol{x}}}

\renewcommand{\a}{{\mathsf a}}
\newcommand{\T}{{\mathsf T}}
\newcommand{\alg}{{\cal{A}}}
\newcommand{\R}{{\cal{R}}}
\newcommand{\ii}{\mathsf{i}}

\begin{document}

\vglue 2cm

\begin{center}
\title{Quantum geometry of 3-dimensional lattices}

\author{
 Vladimir V. Bazhanov,
 Vladimir V. Mangazeev and
 Sergey M. Sergeev}
\address{Department of Theoretical Physics,\\
         Research School of Physical Sciences and Engineering,\\
    Australian National University, Canberra, ACT 0200, Australia.}

\end{center}

\begin{abstract}
We study geometric consistency relations between angles on
3-dimensional (3D) circular quadrilateral lattices --- lattices whose
faces are planar quadrilaterals inscribable into a circle. We show
that these relations generate canonical transformations of a
remarkable ``ultra-local'' Poisson bracket algebra defined on
discrete 2D surfaces consisting of circular quadrilaterals.
Quantization of this structure leads to new solutions of the
tetrahedron equation (the 3D analog of the Yang-Baxter
equation). These solutions generate an infinite number of
non-trivial solutions of the Yang-Baxter equation and also define
integrable 3D models of statistical mechanics and quantum field theory.
The latter can be thought of as describing quantum fluctuations of
lattice geometry. The classical geometry of the 3D
circular lattices arises as a
stationary configuration giving the leading contribution to the
partition function in the quasi-classical limit.

\end{abstract}

\newpage

\vglue 4.8cm
\section{Introduction}

Currently the quantum integrability is mostly understood as
a {\em purely algebraic} phenomenon.
It stems from the Yang-Baxter equation \cite{Yang:1967,Bax72} and
other algebraic structures such as the
affine quantum groups \cite{Dri87,Jim85}
(also called the quantized Kac-Moody algebras), the Virasoro algebra
\cite{BPZ84} and their representation theory.
It is, therefore, quite interesting to learn that some
integrable models of statistical mechanics and
quantum field theory arise also in quantization of models of discrete
differential geometry.
Recently, we have shown \cite{BMS07a,BMS07b} that the
two-dimensional integrable model associated with the
Faddeev-Volkov solution \cite{Volkov:1992,FV:1993,Faddeev:1994}
of the Yang-Baxter equation describes quantum
fluctuations of circle patterns \cite{BSp} connected with the
Thurston's discrete analogues of the conformal transformations of
the two-dimensional plane \cite{St1}.
The Faddeev-Volkov model is an Ising-type model
with continuous spin variables, which are interpreted as fluctuating
radii of the circles.  It
contains a free parameter, which can be identified with
the Planck constant (in the language of Euclidean quantum field
theory) or with the temperature (in the language of statistical mechanics).
The classical geometry of the circle patterns is described by
stationary configurations giving the leading contribution to
 the partition function in the
quasi-classical (or the zero-temperature) limit.

In this paper we consider similar connections between integrable {\em
three-dimensional} (3D) quantum systems and
integrable classical models of 3D discrete differential geometry.
The analog of the Yang-Baxter equation for integrable quantum  systems
in 3D is called the {\em tetrahedron equation}.
It was introduced by Zamolodchikov in
\cite{Zamolodchikov:1980rus,Zamolodchikov:1981kf}  (see also
\cite{Baxter:1986phd,Bazhanov:1992jq,
Bazhanov:1993j,Kashaev:1993ijmp,Korepanov:1993jsp,
KashaevKorepanovSergeev,BS05} for further
results in this field, used in this paper).
Similarly to the Yang-Baxter equation the tetrahedron
equation provides local integrability conditions which are not
related to the size of the lattice. Therefore the same solution of
the tetrahedron equation defines different integrable models on
lattices of different size, e.g., for finite
periodic cubic lattices. Obviously, any such three-dimensional
model can be viewed as a two-dimensional integrable model on a
square lattice, where the additional third dimension is treated as
an internal degree of freedom. Therefore every solution of the
tetrahedron equation provides an infinite sequence of integrable
2D models differing by the size of this ``hidden third
dimension''. Then a natural question arises whether known 2D
integrable models can be obtained in this way.
A complete answer to
this question is yet unknown. So far only
two different (but related) examples of such correspondence
have been constructed. The first example,
connected with the generalized  chiral Potts model
\cite{BKMS,Date:1990bs}, was found in \cite{Bazhanov:1992jq}.
The second example was
recently found in \cite{BS05}.
The corresponding solution of the tetrahedron equation constructed in
\cite{BS05}
reproduces all two-dimensional solvable models related to
finite-dimensional highest weight representations for all
quantized affine algebras $U_q(\widehat{sl}_n)$,
$n=2,3,\ldots,\infty$, where the rank $n$ coincides with the size
of the hidden dimension.
Here we unravel yet another remarkable property of
the same solution of the tetrahedron equation. We show that it
can be obtained from quantization of geometric
integrability conditions for the 3D {\em circular 
lattices} --- lattices whose faces are planar quadrilaterals 
inscribable into a circle.

The 3D circular lattices were introduced \cite{Bob96} as a
discretization of orthogonal coordinate systems, originating from classical
works of Lam\'e \cite{Lame1859} and Darboux \cite{Darboux}.
In the continuous case such coordinate systems are described by
integrable partial differential equations (they are connected with the
classical soliton theory \cite{ZakharovManakov,Kri96}). Likewise, the
quadrilateral and circular lattices are described by integrable 
difference equations. The key idea of the geometric approach
\cite{BobenkoPinkall,Bob96,DoliwaSantini,AdlerBobenkoSuris,
KonopelchenkoSchief,CDS97,KS98, DMS98, BoSurUMN07} 
to integrability of discrete classical systems is to
utilize various consistency conditions 
\cite{BoSur05} arising from geometric relations
between elements of the lattice. It is quite remarkable that these conditions
ultimately reduce to certain 
incidence theorems of elementary geometry. 
For instance, the integrability conditions
for the quadrilateral lattices merely reflect the 
fact of existence of the 4D Euclidean cube \cite{DoliwaSantini}. 
Here we present these conditions algebraically 
in a standard form of the {\em functional tetrahedron equation}
\cite{KashaevKorepanovSergeev}. The latter serves 
as the classical analog of the quantum tetrahedron equation, discussed
above, and
provides a connecting link to integrable quantum systems. 

We study 
relations between edge angles on the 3D circular quadrilateral
lattices and show that these relations describe  symplectic transformations of
a remarkable ``ultra-local'' Poisson algebra on quadrilateral
surfaces (see Eq.\eqref{poisson}).
In Section~\ref{poisson-sec}\
we formulate a variational principle which generates these
angle relations and explicitly calculate the Lagrangian form of the action
functional ${\cal S}^{(cl)}$ (curiously, it expressed through the
Lobachevsky function).
Next, in Section~\ref{quantization}\  we quantize this structure
and obtain two solutions of the tetrahedron equation.
One of them was previously known \cite{BS05}, but another
one, given in Section~\ref{modular-sec} and \ref{IRC},
is new. The solutions are used to define
two different integrable models of statistical mechanics and quantum
field theory on the cubic lattice.
Their partition functions depend on the quantum parameter $\hbar$ (the
Planck constant). For both of these models
the quasi-classical limit of the partition function
\be
Z\simeq \exp\Big({-\frac{{\cal S}^{(cl)}}{\hbar}}\Big),\qquad \hbar\to
0\ ,
\ee
is governed by the above classical action ${\cal S}^{(cl)}$,
determined by the angle geometry of the circular lattices (more
precisely, the two models lead to two different regimes of the
classical action connected by an analytic continuation).
In this paper we only state our main results, the details will be
presented elsewhere \cite{BMS08}.

\section{Discrete differential geometry: ``Existence as integrability''}
In this section we consider classical discrete integrable systems
associated with the quadrilateral lattices.
There are several ways to extract algebraic integrable systems
from the geometry of these lattices. One approach,
developed in
\cite{Kas96,BogdanovKonopelchenko, DoliwaSantini, DS00,
KonopelchenkoSchief},
leads to discrete analogs of the
Kadomtsev-Petviashvili integrable hierarchy.
Here we present a different approach exploiting the angle geometry
of the 3D quadrilateral lattices.

\subsection{Quadrilateral lattices}
Consider three-dimensional lattices, obtained by
embeddings of the integer cubic lattice ${\mathbb Z}^3$ into the
$N$-dimensional Euclidean space ${\mathbb R}^N$, with $N\ge 3$.
Let $\x({m})\in {\mathbb R}^N$, denote coordinates of the lattice
vertices, labeled by the 3-dimensional integer vector
 $m=m_1e_1+m_2e_2+m_3e_3\in
{\mathbb Z}^3$, where $e_1=(1,0,0), e_2=(0,1,0)$ and
$e_3=(0,0,1)$. Further, for any given lattice vertex $\x_0=\x(m)$,
the symbols $\x_i=\x(m+e_i)$, $\x_{ij}=\x(m+e_i+e_j)$, etc., will
denote neighboring lattice vertices. The lattice is called
{\em quadrilateral\/}\  if all its faces $(\x_0,\x_i,\x_j,\x_{ij})$
are planar quadrilaterals.
The existence of these lattices is based on the
following elementary geometry fact (see Fig.~1) \cite{DoliwaSantini},
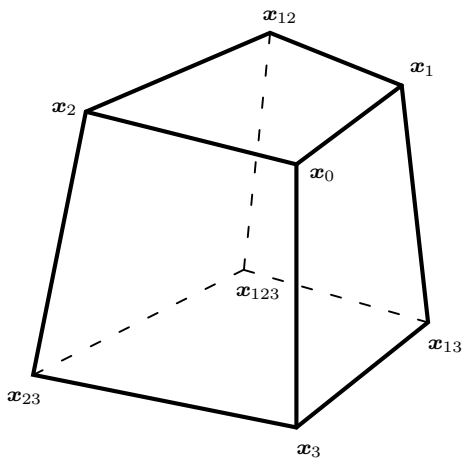
\begin{figure}[ht]
\centering
\setlength{\unitlength}{0.35mm}
\begin{picture}(170,170)
\allinethickness{.6mm}
 \path(0,20)(100,0)(150,40)(140,130)(90,150)(20,120)(0,20)
 \path(20,120)(100,100)
 \path(100,0)(100,100)
 \path(140,130)(100,100)
\allinethickness{.3mm}
 \dashline{5}(0,20)(80,60)
 \dashline{5}(150,40)(80,60)
 \dashline{5}(90,150)(80,60)
 \put(-10,10){\scriptsize $\boldsymbol{x}_{23}$}
 \put(100,-10){\scriptsize $\boldsymbol{x}_{3}$}
 \put(150,30){\scriptsize $\boldsymbol{x}_{13}$}
 \put(143,135){\scriptsize $\boldsymbol{x}_{1}$}
 \put(87,155){\scriptsize $\boldsymbol{x}_{12}$}
 \put(7,120){\scriptsize $\boldsymbol{x}_{2}$}
 \put(77,50){\scriptsize $\boldsymbol{x}_{123}$}
 \put(105,95){\scriptsize $\boldsymbol{x}_{0}$}
%

\end{picture}
\caption{An elementary hexahedron of a cubic quadrilateral lattice.}
\label{fig-cube1}
\end{figure}
\vspace{.2cm}

{\em Consider four points $\x_0, \x_1, \x_2, \x_3$
in general position in ${\mathbb R}^N$, $N\ge 3$.
On each of the three planes $(\x_0,\x_i,\x_j)$,
$1\le i <j\le 3$ choose an extra point $\x_{ij}$ not lying on the lines
$(\x_0,\x_i)$, $(\x_0,\x_j)$ and $(\x_i,\x_j)$. Then there exist a unique point
$\x_{123}$ which simultaneously belongs to the three planes
$(\x_1,\x_{12},\x_{13})$, $(\x_2,\x_{12},\x_{23})$ and
$(\x_3,\x_{13},\x_{23})$.}
\vspace{.2cm}

\noindent The six planes, referred to above, obviously lie in the
same 3D subspace of the target space. They define a hexahedron
with quadrilateral faces, shown in Fig.~\ref{fig-cube1}. It has
the topology of the cube, so we will call it ``cube'', for
brevity. Let us study elementary geometry relations among the
\begin{figure}[ht]
\centering
\setlength{\unitlength}{0.35mm}
\begin{picture}(400,170)
\put(230,10){\begin{picture}(150,150)
\allinethickness{.6mm}
 \path(0,20)(100,0)(150,40)(140,130)(90,150)(20,120)(0,20)
\thinlines
 \dashline{5}(0,20)(80,60)
 \dashline{5}(150,40)(80,60)
 \dashline{5}(90,150)(80,60)
 \put(-10,10){\scriptsize $\boldsymbol{x}_{23}$}
 \put(100,-10){\scriptsize $\boldsymbol{x}_{3}$}
 \put(150,30){\scriptsize $\boldsymbol{x}_{13}$}
 \put(143,135){\scriptsize $\boldsymbol{x}_{1}$}
 \put(87,155){\scriptsize $\boldsymbol{x}_{12}$}
 \put(7,120){\scriptsize $\boldsymbol{x}_{2}$}
 \put(17,21){\scriptsize $\beta_3'$}
 \put(7,32){\scriptsize $\delta_2'$}
 \put(70,64){\scriptsize $\gamma_2'$}
 \put(80,50){\scriptsize $\alpha_3'$}
 \put(85,65){\scriptsize $\delta_1'$}
 \put(22,112){\scriptsize $\beta_2'$}
 \put(78,138){\scriptsize $\alpha_2'$}
 \put(92,138){\scriptsize $\beta_1'$}
 \put(93,5){\scriptsize $\delta_3'$}
 \put(132,36){\scriptsize $\gamma_3'$}
 \put(137,47){\scriptsize $\gamma_1'$}
 \put(131,123){\scriptsize $\alpha_1'$}
 \put(-10,150){(b)}
\end{picture}}
\put(10,10){\begin{picture}(150,150)
\allinethickness{.6mm}
 \path(0,20)(100,0)(150,40)(140,130)(90,150)(20,120)(0,20)
 \path(20,120)(100,100)
 \path(100,0)(100,100)
 \path(140,130)(100,100)
\thinlines
 \put(-10,10){\scriptsize $\boldsymbol{x}_{23}$}
 \put(100,-10){\scriptsize $\boldsymbol{x}_{3}$}
 \put(150,30){\scriptsize $\boldsymbol{x}_{13}$}
 \put(143,135){\scriptsize $\boldsymbol{x}_{1}$}
 \put(87,155){\scriptsize $\boldsymbol{x}_{12}$}
 \put(7,120){\scriptsize $\boldsymbol{x}_{2}$}
 \put(5,23){\scriptsize $\delta_1$}
 \put(22,110){\scriptsize $\beta_1$}
 \put(90,7){\scriptsize $\gamma_1$}
 \put(90,93){\scriptsize $\alpha_1$}
 \put(103,13){\scriptsize $\delta_2$}
 \put(139,42){\scriptsize $\gamma_2$}
 \put(131,117){\scriptsize $\alpha_2$}
 \put(105,95){\scriptsize $\beta_2$}
 \put(95,107){\scriptsize $\delta_3$}
 \put(40,120){\scriptsize $\beta_3$}
 \put(123,127){\scriptsize $\gamma_3$}
 \put(86,141){\scriptsize $\alpha_3$}
 \put(-10,150){(a)}
\end{picture}}
\end{picture}
\caption{The ``front'' (a)  and ``back'' (b) faces of the cube in
Fig.~\ref{fig-cube1} and their angles.}\label{fig-cube2}
\end{figure}
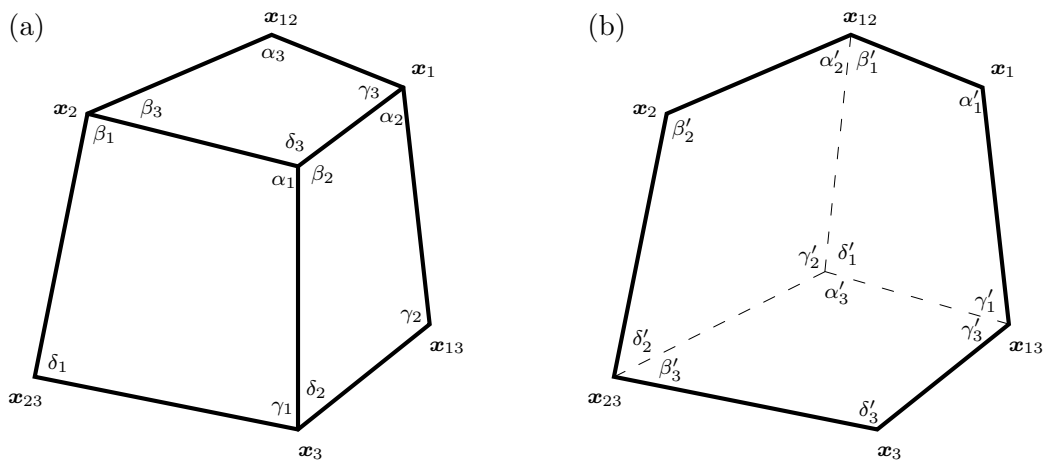
angles of this cube. Denote the angles between the edges as in
Fig.~\ref{fig-cube2}. Altogether we have $6 \times 4=24$ angles,
connected by six linear relations
\begin{equation}
\alpha_j+\beta_j+\gamma_j+\delta_j=2\pi ,\qquad
\alpha'_j+\beta'_j+\gamma'_j+\delta'_j=2\pi, \qquad
j=1,2,3,\label{2pi}
\end{equation}
which can be immediately solved for all ``$\delta$'s''. This
leaves 18 angles, but only nine of them are independent. 
Indeed, a mutual arrangement (up to an overall rotation)
of unit normal vectors to six planes 
in the 3D-space is determined by
nine angles only. 
Once this arrangement is fixed all other angles
can be calculated. Thus the nine independent angles of the three
``front'' faces of the cube, shown in Fig.\ref{fig-cube2}a,
completely determine the angles on the three ``back'' faces, shown
in Fig.\ref{fig-cube2}b, and vise versa. So the geometry of our
cube provides an invertible map for three triples of independent
variables
\begin{equation}
{\cal R}_{123}:\qquad \{\alpha_j,\beta_j,\gamma_j\}\to
\{\alpha'_j,\beta'_j,\gamma'_j\}, \qquad j=1,2,3.\label{map}
\end{equation}
Suppose now that all angles are known. To completely define the
cube one also needs  to specify lengths of its three edges. All
the remaining edges can be then determined from  simple linear
relations. Indeed, the four sides of every quadrilateral are
constrained by two relations, which can be conveniently presented
in the matrix form
\begin{equation}
\left(\begin{array}{c} \ell_p'\\[.1cm]
\ell_q'\end{array}\right)\;=\;
X\,
\left(\begin{array}{c} \ell_p\\[.1cm]
\ell_q\end{array}\right), \qquad
X=
\left(\begin{array}{cc} A(\alg)& B(\alg)\\[.3cm]
C(\alg)&D(\alg)\end{array}\right)
=
\left(\begin{array}{cc} \frac{\sin\gamma}{\sin\delta} &
\frac{\sin(\delta+\beta)}{\sin\delta}\\[.3cm]
\frac{\sin(\delta+\gamma)}{\sin\delta} &
\frac{\sin\beta}{\sin\delta}\end{array}\right) \;
\label{lin-pr}
\end{equation}
where $\alg=\{\alpha,\beta,\gamma,\delta\}$ denotes the set of angles
and $\ell_p,\ell_q,\ell_{p'},\ell_{q'}$  denote the edge lengths,
arranged as in Fig.\ref{fig-kvadrat2}. Note that due to
\eqref{2pi} the entries of the two by two matrix in \eqref{lin-pr}
satisfy the relation
\begin{equation}
AD-BC=(AB-CD)/(DB-AC).\label{constraint}
\end{equation}

Assume that the lengths $\ell_p, \ell_q$, $\ell_r$, on one
side of the two pictures in Fig.\ref{fig-cube3} are given. Let us
find the other three lengths $\ell_{p'}, \ell_{q'}$,
$\ell_{r'}$ on their opposite side, by iterating the relation
\eqref{lin-pr}. Obviously, this can be done in two different ways:
either using the front three faces, or the back ones --- the results
must be the same. This is exactly where the geometry gets into play. The
results must be consistent due to the very existence of the cube
in Fig.~\ref{fig-cube1} as a geometric body.
\begin{figure}[ht]
\centering
\setlength{\unitlength}{0.35mm}
\begin{picture}(100,100)
\allinethickness{.6mm}
 \path(0,10)(80,20)(90,75)(20,80)(0,10)
 \put(-5,0){\scriptsize $\boldsymbol{x}_{23}$}
 \put(80,10){\scriptsize $\boldsymbol{x}_3$}
 \put(15,85){\scriptsize $\boldsymbol{x}_2$}
 \put(90,80){\scriptsize $\boldsymbol{x}_0$}
 \put(8,15){\scriptsize $\delta$}
 \put(72,25){\scriptsize $\gamma$}
 \put(22,70){\scriptsize $\beta$}
 \put(80,67){\scriptsize $\alpha$}
 \thinlines
 \spline(-5,40)(50,50)(100,50)\path(90,52)(100,50)(90,48)
 \put(-15,40){\scriptsize $p$}
 \spline(40,0)(50,55)(60,100)\path(55,92)(60,100)(61,90)
 \put(37,-10){\scriptsize $q$}
 \put(45,5){\scriptsize $\ell_q'$}
 \put(47,83){\scriptsize $\ell_q$}
 \put(0,50){\scriptsize $\ell_p'$}
 \put(90,40){\scriptsize $\ell_p$}
\end{picture}
\caption{The angles $\alg=\{\alpha,\beta,\gamma,\delta\}$ and sides
$\ell_p,\ell_q,\ell_{p'},\ell_{q'}$ of a quadrilateral and the oriented
  rapidity lines.}
\label{fig-kvadrat2}
\end{figure}
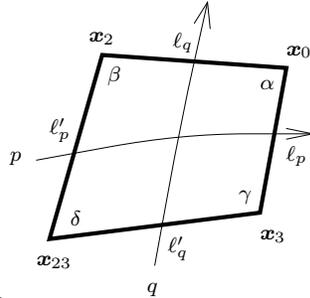
However, they will be consistent only if all geometric relations
between the two sets of angles in the front and back faces of the
cube are taken into account. To write these relations in a
convenient form we need to introduce additional notations. Note,
that Fig.\ref{fig-kvadrat2} shows two thin lines, labeled by the
symbols ``$p$'' and ``$q$''. Each line crosses a pairs of opposite
edges, which we call ``corresponding'' (in the sense that they
correspond to the same thin line). Eq.\eqref{lin-pr} relates the
lengths $(\ell_p,\ell_q)$ of two adjacent edges with the
corresponding lengths $(\ell'_p,\ell'_q)$ on the opposite side of
the quadrilateral.

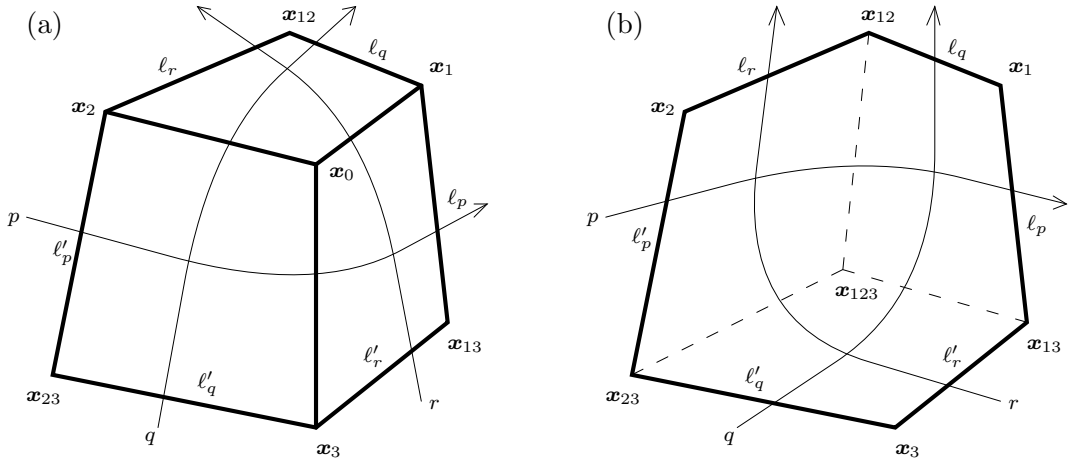
\begin{figure}[ht]
\centering
\setlength{\unitlength}{0.35mm}
\begin{picture}(400,170)
\put(230,10){\begin{picture}(150,150)
\allinethickness{.6mm}
 \path(0,20)(100,0)(150,40)(140,130)(90,150)(20,120)(0,20)
\thinlines
 \dashline{5}(0,20)(80,60)
 \dashline{5}(150,40)(80,60)
 \dashline{5}(90,150)(80,60)
 \put(-10,10){\scriptsize $\boldsymbol{x}_{23}$}
 \put(100,-10){\scriptsize $\boldsymbol{x}_{3}$}
 \put(150,30){\scriptsize $\boldsymbol{x}_{13}$}
 \put(143,135){\scriptsize $\boldsymbol{x}_{1}$}
 \put(87,155){\scriptsize $\boldsymbol{x}_{12}$}
 \put(7,120){\scriptsize $\boldsymbol{x}_{2}$}
 \put(77,50){\scriptsize $\boldsymbol{x}_{123}$}
%
 \spline(-10,80)(85,105)(165,85)\path(160,88)(165,85)(160,83)
 \spline(40,0)(115,50)(115,160)\path(112,153)(115,160)(117,153)
 \spline(140,10)(40,40)(55,160)\path(52,153)(55,160)(58,153)
 \put(-17,78){\scriptsize $p$}
 \put(35,-5){\scriptsize $q$}
 \put(143,7){\scriptsize $r$}
 \put(150,75){\scriptsize $\ell_p$}
 \put(120,142){\scriptsize $\ell_q$}
 \put(40,135){\scriptsize $\ell_r$}
 \put(118,25){\scriptsize $\ell_r'$}
 \put(43,17){\scriptsize $\ell_q'$}
 \put(0,70){\scriptsize $\ell_p'$}
 \put(-10,150){(b)}
\end{picture}}
\put(10,10){\begin{picture}(150,150)
\allinethickness{.6mm}
 \path(0,20)(100,0)(150,40)(140,130)(90,150)(20,120)(0,20)
 \path(20,120)(100,100)
 \path(100,0)(100,100)
 \path(140,130)(100,100)
\thinlines
 \put(-10,10){\scriptsize $\boldsymbol{x}_{23}$}
 \put(100,-10){\scriptsize $\boldsymbol{x}_{3}$}
 \put(150,30){\scriptsize $\boldsymbol{x}_{13}$}
 \put(143,135){\scriptsize $\boldsymbol{x}_{1}$}
 \put(87,155){\scriptsize $\boldsymbol{x}_{12}$}
 \put(7,120){\scriptsize $\boldsymbol{x}_{2}$}
 \put(105,95){\scriptsize $\boldsymbol{x}_{0}$}
 \spline(-10,80)(100,50)(165,85)\path(160,85)(165,85)(160,80)
 \spline(40,0)(60,110)(115,160)\path(110,159)(115,160)(113,153)
 \spline(140,10)(120,115)(55,160)\path(58,155)(55,160)(60,159)
 \put(-17,78){\scriptsize $p$}
 \put(35,-5){\scriptsize $q$}
 \put(143,7){\scriptsize $r$}
 \put(150,85){\scriptsize $\ell_p$}
 \put(120,142){\scriptsize $\ell_q$}
 \put(40,135){\scriptsize $\ell_r$}
 \put(118,25){\scriptsize $\ell_r'$}
 \put(55,15){\scriptsize $\ell_q'$}
 \put(0,65){\scriptsize $\ell_p'$}
 \put(-10,150){(a)}
\end{picture}}
\end{picture}
\caption{The ``front'' (a) and ``back'' (b) faces of the cube in
Fig.~\ref{fig-cube1} and ``rapidity'' lines.}\label{fig-cube3}
\end{figure}

Consider now Fig.\ref{fig-cube3}a which contains three directed thin lines
connecting corresponding edges of the three quadrilateral faces.
By the analogy with
the 2D Yang-Baxter equation, where similar arrangements
occur, we call them ``rapidity''
lines\footnote{%
However, at the moment we do not assume any further meaning for
these lines apart from using them as a convenient way of labeling
to the corresponding (opposite) edges of quadrilaterals.}. We will
now apply \eqref{lin-pr} three times starting from the top face
and moving against the directions of the arrows.  Introduce the
following three by three matrices
\begin{equation}
X_{pq}(\alg)=\left(\begin{array}{ccc} A& B
& 0 \\ C& D& 0 \\
0 & 0 & 1\end{array}\right)\;,\quad
X_{pr}(\alg)=\left(\begin{array}{ccc} A & 0 &
    B
\\ 0 & 1 & 0
\\ C& 0 &  D\end{array}\right)\;,
\quad
X_{qr}(\alg)=\left(\begin{array}{ccc} 1 & 0 & 0 \\ 0 & A& B \\
0 & C& D\end{array}\right)\;,
\end{equation}
where $A,B,C,D$ are defined in \eqref{lin-pr} and their dependence
on the angles $\alg=\{\alpha,\beta,\gamma,\delta\}$ is implicitly
understood. It follows that
\begin{equation}
  (\ell'_p,\ell'_q,\ell'_r)^t\,=\,X_{pq}(\alg_1)\,X_{pr}(\alg_2)
  \,X_{qr}(\alg_3)\ (\ell_p,\ell_q,\ell_r)^t
\end{equation} where
\begin{equation}
\alg_j=\{\alpha_j,\beta_j,\gamma_j,\delta_j\},\quad j=1,2,3,
\label{A-def}
\end{equation}
the lengths $\ell_p,\ell_q,\ldots$ are defined as in
Fig.\ref{fig-cube3}, and the superscript ``$t$'' denotes the
matrix transposition. Performing  similar calculations for the
back faces in Fig.\ref{fig-cube3}b and equating the resulting
three by three matrices, one obtains
\begin{equation}
X_{pq}(\alg_1)\,X_{pr}(\alg_2) \,X_{qr}(\alg_3)\,=\,
X_{qr}(\alg'_3)\, X_{pr}(\alg'_2)\, X_{pq}(\alg'_1)\ .\label{lybe}
\end{equation}
where
\begin{equation}
\alg'_j=\{\alpha'_j,\beta'_j,\gamma'_j,\delta'_j\},\quad j=1,2,3\ .
\label{Ap-def}
\end{equation}
This matrix relation contains exactly nine scalar equations where the 
LHS only depends on the front angles \eqref{A-def}, while the RHS only
depends on the back angles \eqref{Ap-def}. Solving these equations
one can obtain explicit form of the map \eqref{map}. The resulting
expressions are rather complicated and not particularly useful.
However the mere fact that the map \eqref{map} satisfy
a very special Eq.\eqref{lybe} is extremely important. Indeed, rewrite this
equation as
\begin{equation}
X_{pq}(\alg_1)\,X_{pr}(\alg_2) \,X_{qr}(\alg_3)\,=\,\R_{123}\,
\Big( X_{qr}(\alg_3)\, X_{pr}(\alg_2)\, X_{pq}(\alg_1)\Big)
\label{tza}
\end{equation}
where $\R_{123}$ is an operator acting as the  substitution
\eqref{map} for any function
$F(\alg_1,\alg_2,\alg_3)$ of the angles,
\begin{equation}
\R_{123}\Big(F(\alg_1,\alg_2,\alg_3)\Big)=F(\alg_1',\alg_2',\alg_3')
\end{equation}
Then, following the arguments of \cite{Korepanov:1993jsp}, one can
show that the map \eqref{map} satisfies the {\em functional
tetrahedron equation} \cite{KashaevKorepanovSergeev}
\begin{equation}
\R_{123} \cdot\R_{145}\cdot \R_{246} \cdot\R_{356} \;=\;
\R_{356}\cdot \R_{246} \cdot\R_{145} \cdot\R_{123}\label{fte}\ ,
\end{equation}
where both sides are
compositions of the maps \eqref{map}, involving six different sets
of angles. Algebraically, this equation arises
as an associativity condition for the
cubic algebra \eqref{tza}.
To discuss its geometric meaning
we need to introduce {\em discrete evolution systems} associated
with the map \eqref{map}.

\subsection{Discrete evolution systems: ``Existence as
  integrability''}

\begin{figure}[htb]
\centering
\includegraphics[scale=0.62]{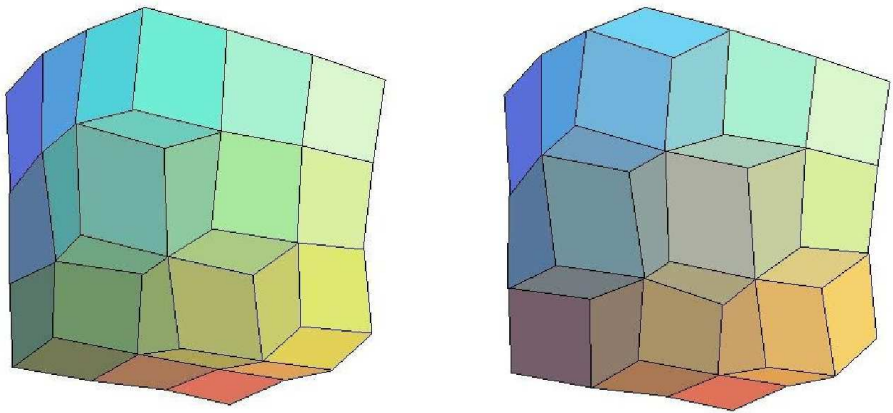}
\caption{Visualization of the 3D ``staircase'' evolution.}\label{fig-viz}
\end{figure}

Consider a sub-lattice $L$ of the 3D quadrilateral lattice,
which only includes points $\x(m)$ with
$m_1,m_2,m_3\ge0$.  The boundary of this sub-lattice is a 2D discrete
surface formed by quadrilaterals with the vertices
$\x(m)$ having at least one of their integer
coordinates $m_1,m_2,m_3$ equal to zero and the other two
non-negative.
Assume that all quadrilateral angles on this surface are known, and
consider them as initial data. Then repeatedly applying the
map \eqref{map} one can calculate angles on all faces of the
sub-lattice $L$, defined above (one has to start from the corner
${\x}(0)$).
The process can be visualized as an evolution of
the initial data surface where
every transformation \eqref{map} corresponds to
a ``flip'' between the front and back faces (Fig.~\ref{fig-cube2})
of some cube adjacent to the surface.
This makes the
surface looking as a 3D ``staircase'' (or a pile of cubes) in the
intersection corner of the three coordinate planes,
see Fig.~\ref{fig-viz} showing  two stages of this process.
Note, that the corresponding evolution equations can be written in a
covariant form for an arbitrary lattice cube (see Eq.\eqref{cov-form}
below for an example). It is also useful to have in mind that the
above evolution can be defined purely geometrically as a
{\em ruler-and-compass\/} type construction. Indeed the construction of the
point ${\x}_{123}$ in Fig.~\ref{fig-cube1} from the points
$\x_0,\x_1,\x_2,\x_3,\x_{12},\x_{13},\x_{23}$
(and that is what is necessary for flipping a cube)
only requires a {\em 2D-ruler\/} which allows to draw planes through
any three non-collinear points in the Euclidean space.

\begin{figure}[ht]
\centering
\includegraphics[scale=0.60]{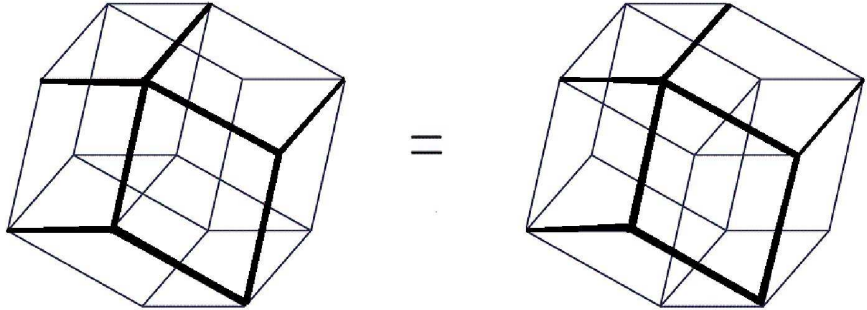}
\caption{Two dissections of the rhombic dodecahedron into four
quadrilateral hexahedra.} \label{fig-tetra}
\end{figure}

Similar evolution systems can be defined for other
quadrilateral lattices instead
of the 3D cubic lattice considered above.
Since the evolution is local (only one cube is flipped at a
time) one could consider finite lattices as well. For example, consider
six adjacent quadrilateral faces covering the front surface of the
rhombic dodecahedron\footnote{%
It is worth noting that the most general rhombic dodecahedron
with quadrilateral faces can only be embedded into (at least)
the 4D Euclidean space.}
shown in Fig.~\ref{fig-tetra}. Suppose that all  angles on
these faces are given and consider them as initial data.
Now apply a sequence of four maps \eqref{map} and calculate angles on
the back surface of the rhombic dodecahedron.
This can be done in two alternative ways, corresponding to the two
different dissection of the rhombic dodecahedron into four cubes shown
in Fig.\ref{fig-tetra}.
The functional tetrahedron equation \eqref{fte} states
that the results will be the same. Thereby it gives an algebraic proof for
the equivalence of two ``ruler-and-compass'' type constructions of the
back surface of the dodecahedron in Fig.~\ref{fig-tetra}.
Can we also prove this equivalence geometrically?
Although from the first sight this
does not look trivial, it could be easily done from the point of view of
the 4D geometry.
The required statement follows just from the fact of existence of the
quadrilateral lattice with the topology of the 4D cube
\cite{DoliwaSantini}. The latter is
defined by eight intersecting 3-planes in a general position in the 4-space.
The two rhombic dodecahedra shown in Fig.~\ref{fig-tetra}
are obtained by a dissection of the 3-surface of this 4-cube, along its
2-faces, so these dodecahedra
must have exactly the same quadrilateral 2-surface.
Thus the functional tetrahedron equation \eqref{fte}, which plays the
role of integrability condition for the discrete evolution system
associated with the map \eqref{map}, simply follows from the mere fact of
existence of the 4-cube, which is the simplest 4D quadrilateral lattice.
For a further discussion of a relationship between the geometric
consistency and integrability see \cite{BoSur05}.

Note that, to our knowledge, the linear problem \eqref{lin-pr} for
the lengths of the quadrilateral with coefficients depending on
the angles was not hitherto considered. Much attention was given
to the linear problem for coordinates of lattice vertices
introduced in \cite{DoliwaSantini}. In the notations of
Fig.~\ref{fig-kvadrat2} it reads
\begin{equation}
\boldsymbol{x}_{23}-\boldsymbol{x}_0 \;=\;
Y_{23}
\;(\boldsymbol{x}_2-\boldsymbol{x}_0) \;+\;
Y_{32}
\;(\boldsymbol{x}_3-\boldsymbol{x}_0),\qquad
Y_{23}=\frac{\sin\beta}{\sin\alpha}\frac{\ell_p'}{\ell_p},\qquad
Y_{32}=\frac{\sin\gamma}{\sin\alpha} \frac{\ell_q'}{\ell_q}\ ,
\end{equation}
where ${\x}_0,{\x}_{2},{\x}_{3},{\x}_{23}$
 are 3D coordinate of the vertices. The
evolution equations \cite{BogdanovKonopelchenko, DoliwaSantini} for
 the coefficients $Y_{ij}$ mix
 lengths and angles in a complicated way . To the contrary, our map
\eqref{map} only involve angle variables. It describes
internal geometry of the quadrilateral lattice independently of
its embedding into the target space (the latter, of course,
depends on the length variables).

Finally, note that the same map \eqref{map}
also arises if one considers a linear problem for unit vectors
associated with directions of the lattice edges. These vectors
satisfy the relation (in the notation of Fig.\ref{fig-kvadrat2})
\begin{equation}
\tvec{{\bf v}_{p}'}{{\bf v}_{q}'}\;=\;
\left({X}^{-1}\right)^{t}\tvec{{\bf  v}_{p}^{}}{{\bf v}_{q}^{}}
\end{equation}
where $X$ is the two by two matrix defined in \eqref{lin-pr} and
\begin{equation}
{\bf v}_p^{}=\frac{\x_3-\x_0}{\ell_p}\;,\quad {\bf
v}_q^{}=\frac{\x_0-\x_2}{\ell_q}\;,\quad {\bf
v}_p'=\frac{\x_{23}-\x_2}{\ell_p'}\;,\quad {\bf
v}_q'=\frac{\x_3-\x_{23}}{\ell_q'}\;.
\end{equation}
The property (\ref{constraint}) follows from the condition $ {\bf
v}^2=1$.

\section{Variational principle for the 3D circular
 lattices}\label{poisson-sec}
\subsection{Poisson structure of circular lattices}
\begin{figure}[hbt]
\centering
\includegraphics[scale=.75]{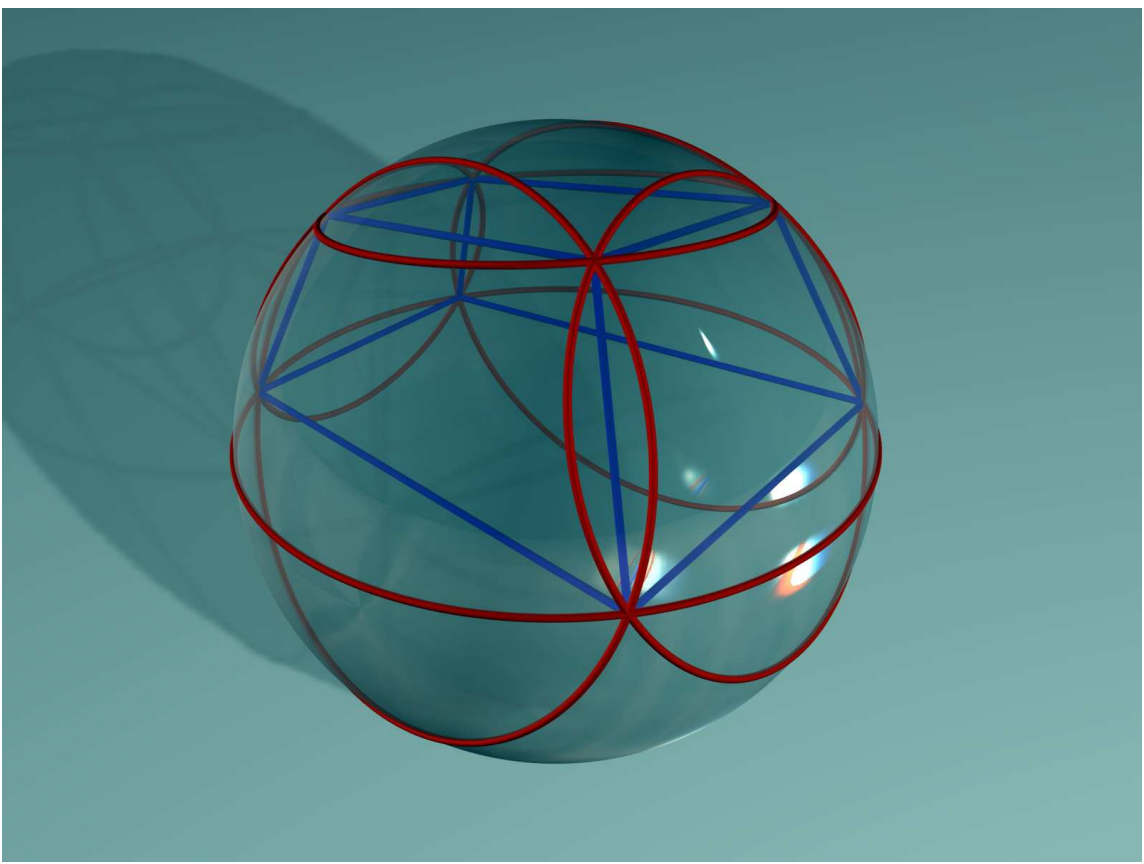}
\caption{Miquel configuration of circles in 3D space, an elementary
hexahedron and its circumsphere.}\label{miguel}
\end{figure}

The 3D {\em circular lattice} \cite{Bob96,CDS97,KS98} is a
special 3D quadrilateral lattice where all faces are circular
quadrilaterals (i.e., quadrilaterals which can be inscribed into a
circle).  The existence of these lattices is established by the
following beautiful geometry theorem due to Miquel \cite{Miquel}
(see Fig.~\ref{miguel})

\noindent {\bf Miquel theorem.} {\em Consider four
points $\x_0, \x_1, \x_2, \x_3$ in general position in ${\mathbb
R}^N$, $N\ge 3$. On each of the three circles $c(\x_0,\x_i,\x_j)$,
$1\le i <j\le 3$ choose an additional new point $\x_{ij}$. Then
there exist a unique point $\x_{123}$ which simultaneously belongs
to the three circles $c(\x_1,\x_{12},\x_{13})$,
$c(\x_2,\x_{12},\x_{23})$ and $c(\x_3,\x_{13},\x_{23})$.}

\vspace{.2cm}
\noindent
It is easy to see that the above six circles lie on the same
sphere. It follows then that every elementary
``cube'' on a circular lattice (whose vertices are
at the circle intersection points) is inscribable into a sphere, see
Fig.~\ref{miguel}.
The general formulae of the previous subsection can be readily
specialized for the circular lattices. A circular quadrilateral has
only two independent angles.
In the notation of Fig.~\ref{fig-kvadrat2} one has
\begin{equation}
\gamma=\pi-\beta,\qquad \delta=\pi-\alpha\ .\label{circ-ang}
\end{equation}
Due to the Miquel theorem we can simply impose these restrictions on
all faces of the lattice without running to any contradictions.
The two by two matrix in \eqref{lin-pr} takes the form
\begin{equation}
X\;=\;\left(\begin{array}{cc} k & a^* \\[.2cm]
                             -a & k
\end{array}\right),
\qquad \det \, X=1\label{detx}\ ,
\end{equation}
where we have introduced new variables
\begin{equation}
k\;=\;(\csc \alpha)\, {\sin\beta}\;,\quad
a=({\csc\alpha})\,{\sin(\alpha+\beta)}\;,\quad
a^*=({\csc\alpha})\,{\sin(\alpha-\beta)},\label{newvar}
\end{equation}
instead of the two angles $\{\alpha,\beta\}$.
Note that the new variables  are constrained by the relation
\be
aa^*=1-k^2\ . \label{aak-rel}
\ee
Conversely, one has
\begin{equation}
\cos\alpha=\frac{a-a^*}{2k}\;,\quad \cos\beta=\frac{a+a^*}{2}\ .
\end{equation}
Let the variables $\{k_j,a_j,a_j^*\}$,
$\{k'_j,a'_j,a^{*\prime}_j\}$, $j=1,2,3$,\   correspond to the front
and back faces of the cube. The map \eqref{map} then read
explicitly
\begin{equation}\label{themap}
{\mathcal R}_{123}:\qquad
\left\{\begin{array}{ll}
(k_2^{}a_1^*)'=\; k_3^{}a_1^* - \varepsilon k_1^{}a_2^*a_3^{}, &
(k_2^{}a_1^{})'=k_3^{}a_1^{}- \varepsilon k_1^{}a_2^{}a_3^*, \\\\
(a_2^*)'  = a_1^*a_3^* + \varepsilon k_1^{}k_3^{}a_2^*, &
(a_2^{})' = a_1^{}a_3^{} + \varepsilon k_1^{}k_3^{}a_2^{}, \\\\
(k_2^{}a_3^*)'  = k_1^{}a_3^* - \varepsilon k_3^{}a_1^{}a_2^*, &
(k_2^{}a_3^{})' = k_1^{}a_3^{} - \varepsilon k_3^{}a_1^*a_2^{},
\end{array}\right.
\end{equation}
where $\varepsilon=+1$ and
\be
k_2^{\prime}=\sqrt{1-a_2'a_2^{*\prime}}%
\ .
\label{aak2}
\ee

At this point we
note that exactly the same map together with the corresponding equations
\eqref{lybe} and \eqref{fte} were previously obtained in \cite{BS05}.
Moreover, it was discovered that this map is a canonical transformation
preserving the Poisson algebra
\begin{equation}\label{o-poisson}
\{a_i,a_j^*\}\;=\;2\,\delta_{ij}\,k_i^2\;, \quad \{k_i,a_j\}=
\delta_{ij}\,k_i\,a_i\;, \quad
\{k_i,a_j^*\}=-\,\delta_{ij}\,k_i\,a_i^*,\qquad i,j=1,2,3\ ,
\end{equation}
where $k_i^2=1-a_ia_i^*$. Note that variables $k,a,a^*$ on different
quadrilaterals are in involution.
The same Poisson algebra in terms of angle variables reads
\begin{equation}\label{poisson}
\{\alpha_i,\beta_j\}= \delta_{ij}\;,\quad
\{\alpha_i,\alpha_j\}=\{\beta_i,\beta_j\}=0\;.
\end{equation}
This  ``ultra-local'' symplectic structure trivially extends to
any circular quad-surface of initial data, discussed above.
To resolve an apparent ambiguity in naming of the angles,
this surface must be equipped with oriented rapidity lines, similar to
those in Fig.~\ref{fig-cube3}\footnote{We refer
  the reader to our previous paper \cite{BMS07a} where the relationship
  between the rapidity graphs and quadrilateral lattices is thoroughly
  discussed.}.
In addition, the angles for each quadrilateral should be arranged as in
Fig.~\ref{fig-kvadrat2}.
Then one can assume
that the indices $i,j\/$\ \ in \eqref{poisson}   refer to all
quadrilaterals on this surface.

Thus,
the evolution defined by the map \eqref{themap} is a symplectic
transformation. The corresponding equations of motion for the whole
lattice (the analog of the Hamilton-Jacobi equations) can be written
in a ``covariant'' form. For every cube define
\begin{equation}
A_{32}^{}=a_1^{},\;\; A_{23}^{}=a_1^*,\;\;
A_{31}^{}=a_2',\;\; A_{13}^{}=a_2^{'*},\;\;
A_{21}^{}=a_3^{},\;\; A_{12}^{}=a_3^*\label{cov-form}
\end{equation}
where $A_{jk}$ stands for $A_{jk}({m})$, where $m$ is such that ${\x}({m})$
coincides with the coordinates of the top front corner of the cube
(vertex ${\x}_0$ in Fig.\ref{fig-cube1}).
Let $T_k$ be the shift operator $T_k\, A_{ij}({m})=A_{ij}({m}+{e}_k)$. Then
\begin{equation}\label{covariant}
\widetilde{T}_k A_{ij} \;=\;
\frac{A_{ij}-A_{ik}A_{kj}}{K_{ik}K_{kj}}\;,\quad
K_{ij}=K_{ji}=\sqrt{1-A_{ij}A_{ji}}\;,
\end{equation}
where $(i,j,k)$ is an arbitrary permutation of $(1,2,3)$ and
\begin{equation}
\widetilde{T}_1^{}=T_1^{}\;,\quad
\widetilde{T}_2^{}=T_2^{-1}\;,\quad \widetilde{T}_3^{}=T_3^{}\ .
\end{equation}
Note that Eq.(\ref{covariant}) also imply
\be
(\widetilde{T}_k K_{ij})K_{kj} = (\widetilde{T}_i K_{kj}) K_{ij}\ .
\label{kk-rel}
\ee

{\bf Remarks}. The equations \eqref{covariant} have been previously obtained 
in \cite{KS98}, see Eq.(7.20) therein. The quantites $A_{ij}$ in
\eqref{covariant} should be identified with the {\em rotation
  coefficients} denoted as $\tilde\beta_{ij}$ in \cite{KS98}.
The same equations \eqref{covariant} are discussed in \S2.2 of
\cite{BoSur05}, where one can also find a detailed bibliography on the
circular lattices (we are indebted to A.I.Bobenko for these
important remarks).

\subsection{Variational principle: the Lagrangian and action.}

Every canonical transformation
\begin{equation}
P_i,Q_i\;\to\; P'_i,Q'_i\;:\qquad \sum_i dP_i\wedge dQ_i = \sum_i 
dP'_i\wedge dQ'_i
\end{equation}
defines a generating function ${\cal L}(Q,Q')$:
\begin{equation}
P_i=\frac{\partial {\cal L}}{\partial Q_i}\;,\quad P'_i=-\frac{\partial
{\cal L}}{\partial Q'_i}\;\;:\qquad d{\cal L}=\sum_i
\Big(P_i\,dQ_i\;-\;P'_i\,dQ'_i\Big)\;.
\end{equation}
For discrete (in time) canonical transformations, this generating
function coincides with the Lagrangian density of the action
functional. There are many ways to choose canonical variables; the
resulting Lagrangians differ by equivalence transformations. A
convenient choice is the variables $\log k_i$ and $\frac{1}{2}\log v_i$,
\begin{equation}
v_i=\frac{a_i^*}{a_i},\qquad \big\{\log k_i\,,\, \log
v_j\big\}=2\,\delta_{ij}\ .
\end{equation}
The Lagrangian density ${\cal L}={\cal L\/}(v,v')$, associated with one cube, 
 is defined as
\begin{equation}\label{dS}
d{\cal L\/}(v,v')\;=\;\frac{1}{2}\,\sum_{j=1}^3 \big(\log k_j \; d\log v_j^{} - \log
k_j^{\prime} \; d\log v_j'\;\big),
\end{equation}
where $v=(v_1,v_2,v_3)$ and similarly for $v'$.
The equations of motion (\ref{themap}) imply the relations
\begin{equation}
\begin{array}{lll}
k_1^2=\frac{\Big(1-\frac{v_2'}{v_1^{}v_3^{}}\Big)
\Big(1-\frac{v_2^{}}{v_1^{}v_3'}\Big)}
{\Big(1-\frac{v_2^{}}{v_1^{}v_3^{}}\Big)
\Big(1-\frac{v_2'}{v_1^{}v_3'}\Big)}\;,&
 k_2^2=\frac{\Big(1-\frac{v_2^{}}{v_1^{}v_3^{}}\Big)
\Big(1-\frac{v_2^{}}{v_1'v_3'}\Big)}
{\Big(1-\frac{v_2^{}}{v_1'v_3^{}}\Big)
\Big(1-\frac{v_2^{}}{v_1^{}v_3'}\Big)}\;,&
 k_3^2=\frac{\Big(1-\frac{v_2'}{v_1^{}v_3^{}}\Big)
\Big(1-\frac{v_2^{}}{v_1'v_3^{}}\Big)}
{\Big(1-\frac{v_2^{}}{v_1^{}v_3^{}}\Big)
\Big(1-\frac{v_2'}{v_1'v_3^{}}\Big)}\;,
\\&&\\
 k_1^{\prime 2} =
\frac{\Big(1-\frac{v_2^{}}{v_1'v_3'}\Big)
\Big(1-\frac{v_2'}{v_1'v_3^{}}\Big)}
{\Big(1-\frac{v_2'}{v_1'v_3'}\Big)
\Big(1-\frac{v_2^{}}{v_1'v_3^{}}\Big)}\;,&
 k_2^{\prime 2} =
\frac{\Big(1-\frac{v_2'}{v_1'v_3'}\Big)
\Big(1-\frac{v_2'}{v_1^{}v_3^{}}\Big)}
{\Big(1-\frac{v_2'}{v_1'v_3^{}}\Big)
\Big(1-\frac{v_2'}{v_1^{}v_3'}\Big)}\;,&
 k_3^{\prime 2} =
\frac{\Big(1-\frac{v_2^{}}{v_1'v_3'}\Big)
\Big(1-\frac{v_2'}{v_1^{}v_3'}\Big)}
{\Big(1-\frac{v_2'}{v_1'v_3'}\Big)
\Big(1-\frac{v_2^{}}{v_1^{}v_3'}\Big)}\;.
\end{array}\label{kv}
\end{equation}
Introduce new variables
\begin{equation}\label{v2E}
\begin{array}{llll}
\ds e^{-2\Omega_2}=\frac{v_2'}{v_1'v_3'},&
\ds e^{-2\Omega_1}=\frac{v_2}{v_1v_3'},&
\ds e^{-2\Omega_0}=\frac{v_2'}{v_1v_3},&
\ds e^{-2\Omega_3}=\frac{v_2}{v_1'v_3},\\
&&&\\
\ds e^{2\Omega_2'}=\frac{v_2}{v_1v_3},&
\ds e^{2\Omega_1'}=\frac{v_2'}{v_1'v_3},&
\ds e^{2\Omega_0'}=\frac{v_2}{v_1'v_3'},&
\ds e^{2\Omega_3'}=\frac{v_2'}{v_1v_3'},\\
\end{array}
\end{equation}
such that
\be
\Omega'_k=\Omega_k-(\Omega_0+\Omega_1+\Omega_2+\Omega_3)/2\label{O-rel},
\qquad k=0,1,2,3\ .
\ee
Remarkably, the differential \eqref{dS} depends on four
independent variables only.
Indeed rewriting $d{\cal L}(v,v')$ \ in terms of six
independent variables $v_1,v_3,\Omega_0,\Omega_1,\Omega_2,\Omega_3$, one can
easily see that all its dependence on $v_1$, $v_3$ drops out,
\begin{equation}
d{\cal L}(\Omega)=\frac{1}{4}\,\sum_{j=0}^3\,\log
\left(\frac{\sinh^2\Omega_j\sinh^2\Omega'_j}
{\sinh\Omega'_0\,\sinh\Omega'_1\,\sinh\Omega'_2\,\sinh\Omega'_3}
\right)\>d\Omega_j\ . \label{dS1}
\end{equation}
Integrating the last equation, one obtains
\be\label{Sv}
{\cal L}(v,v')= \frac{1}{2}\, \sum_{k=0}^3 \Big( \Lambda\big(\/\Omega_k\/\big)
+\Lambda\big(\/\Omega'_k\/\big)\Big)\ ,
\ee
where $\Lambda(x)$ is the modified Lobachevsky function,
\begin{equation}
\Lambda(x)=\int_0^x \log|2\sinh y|\, dy\ .
\end{equation}
The variables $\Omega_k,\Omega_k'$ in the RHS of \eqref{Sv} are understood as
functions of $v,v'$, defined in \eqref{v2E}.

In the above derivations of
the equations of motion \eqref{kv} and the Lagrangian density
\eqref{Sv}, we considered an isolated cube and the associated face variables
$\{k_j,v_j\}$, \ $\{k'_j,v'_j\}$, \ $j=1,2,3$.
Restoring now the coordinates $m\in\mathbb{Z}^3$ for the whole lattice
(see the definitions before Fig.\ref{fig-cube1}),
 \be
\begin{array}{rclrcl}
v_j&\to&v_j^{}(m),\qquad &v_j'&\to &v_j^{}(m+e_j),\\[0.2cm]
k_j&\to&k_j(m),\qquad &k'_j&\to &k_j(m+e_j)
\end{array}
\qquad j=1,2,3
 \ee
one has for the total action
\begin{equation}\label{actv}
{\cal S}^{(cl)}(\{v\})=\sum_{m\in\mathbb{Z}^3}\; {\cal L}(v(m),v'(m))\;.
\end{equation}
where $v'_j(m)=v_j(m+e_j)$.
The variational principle for this action
leads to the Lagrangian equations of motion
with respect to the variables $v_j(m)$. To write them in a covariant
form it is convenient to define new variables
\begin{equation}
u_1(m)=v_1(m)\;,\quad u_2(m)=1/v_2(m)\;, \quad u_3(m)=v_3(m)\ .
\end{equation}
For an arbitrary lattice site $m\in {\mathbb Z}^3$  denote
\be
u_i^{(0)}\equiv u_i(m), \qquad u_i^{(j)}\equiv u_i(m+e_j), \qquad
u_i^{(jk)}\equiv u_i(m+e_j+e_k),
\ee
where the indices $i,j,k$ independently take any of the three values $1,2,3$.
Then the Lagrangian equations of motion,
determining the stationary point of the action \eqref{actv} take the
form
\begin{equation}\label{eight-bra}
\frac{\big(1-u_i^{(i)}u_j^{(ij)}u_k^{(i)}\big)\
\big(1-u_i^{(i)}u_j^{(i)}u_k^{(ik)}\big)}
{\big(1-u_i^{(i)}u_j^{(i)}u_k^{(i)}\big)\
\big(1-u_i^{(i)}u_j^{(ij)}u_k^{(ik)}\big)}=
\frac{\big(1-u_i^{(i)}u_j^{(j)}u_k^{(0)}\big)\
\big(1-u_i^{(i)}u_j^{(0)}u_k^{(k)}\big)}
{\big(1-u_i^{(i)}u_j^{(0)}u_k^{(0)}\big)\
\big(1-u_i^{(i)}u_j^{(j)}u_k^{(k)}\big)}\;,
\end{equation}
where $(i,j,k)$ is any permutation of $(1,2,3)$. Note that similar,
but different, equations of motion arose previously in \cite{KMS1997,KS98}.

It is sometimes convenient to use
the $k$-variables, $k_j(m)$, instead of the $v$-variables.
The corresponding Lagrangian density for one cube
is obtained from \eqref{Sv} with a Legendre transform
 \begin{equation}\label{S-1}
\overline{{\cal L}}(k,k') = -\frac{1}{2}\,
\sum_{j=1}^3 \log k_j^{} \log v_j +
     {\cal L}(v,v') +
\frac{1}{2}\,\sum_{j=1}^3 \log k_j' \log v_j^{\prime}\;.
\end{equation}
where the variables $v,v'$ are now considered as functions of
$k,k'$\ implicitly defined by \eqref{kv}.
The total action
\begin{equation}\label{actk}
{\cal S}^{(cl)}(\{k\})=
\sum_{m\in\mathbb{Z}^3}\; \overline{{\cal L}}(k(m),k'(m))\;.
\end{equation}
obviously coincides with \eqref{actv}, up to boundary terms.
Note that the Lagrangian ${\cal L}(v,v')$ has a gauge symmetry:
it depends only on four combinations of the six independent
variables $v_j,v_j'$ (see (\ref{v2E},\ref{O-rel}) above).
As a result the variables $k_j,k_j'$ in $\overline{\cal
L}(k,k')$ are constrained by two relations
\begin{equation}\label{constraint-k}
k_1^{}k_2^{}=k_1'k_2'\;,\quad k_2^{}k_3^{}=k_2'k_3'\ ,
\end{equation}
which correspond to (\ref{kk-rel}). Here we prefer to avoid
Lagrangian multipliers and deal with the restricted configuration
space directly. The local constraint (\ref{constraint-k}), taken
for all lattice sites, leads to certain restrictions for boundary
values of $k$'s and makes the Lagrangian equations of motion more
complicated than those in the $v$-variables. For instance, the
local equations of motion in the $k$-variables involve eight
adjacent hexahedra, while those in $v$-variables, given by
\eqref{eight-bra}, involve only two. Nevertheless, the Dirichlet
problem for the action (\ref{actk}) is well posed \cite{BMS08}.

Note also that the logarithmic terms in \eqref{S-1} can be written as
\begin{equation}\label{logterm}
-\sum_{j=1}^3 \log k_j^{}\log v_j + \sum_{j=1}^3 \log k_j' \log
v_j' = - \sum_{j=0}^3\Big( \Omega_j\log (2\sinh\Omega_j) +
\Omega_j' \log (2\sinh\Omega_j')\Big)\ .
\end{equation}

\section{Quantization.}\label{quantization}
\subsection{Tetrahedron equation}
In this section we construct two different, but related
quantizations  of the map
\eqref{themap} and obtain two solutions of the full quantum
tetrahedron equation (see Eq.\eqref{TE} below).
In both cases we start with the canonical quantization of the Poisson
algebra \eqref{poisson},
\be
[\alpha_i,\beta_j]=\xi\, \hbar \,\delta_{ij}\,,\qquad
[\alpha_i,\alpha_j]=0\,,\qquad
[\beta_i,\beta_j]=0\,,\label{ang-alg}
\ee
where $\hbar$ is the quantum parameter (the Planck constant) and $\xi$
is a numerical coefficient, introduced for a further convenience. The
indices $i,j$ label the faces of the ``surface of initial data''
discussed above.
Since the commutation relations \eqref{ang-alg} are ultra-local (in
the sense that the angle variables on different faces commute with
each other), let us concentrate on the local Heisenberg algebra,
\be
{\mathsf H}:\qquad\qquad  [\alpha,\beta]=\xi\, \hbar\ ,\label{H-sing}
\ee
for a single lattice face (remind that the angles
shown in Fig.~\ref{fig-kvadrat2} are related by \eqref{circ-ang}).
The map \eqref{themap} contains the quantities $k,a,a^*$,
defined in \eqref{newvar}, which
now become operators. For definiteness, assume that the
non-commuting factors in \eqref{newvar} are ordered exactly as
written. Then the definitions \eqref{newvar} give
\begin{eqnarray}
k\phantom{{}^*}&=&(U-U^{-1})^{-1}\, (V-V^{-1}),\nonumber\\[0.2cm]
a\phantom{{}^*}&=&q^{-\frac{1}{2}}\,(U-U^{-1})^{-1}\,
(U\,V-U^{-1}\,V^{-1}),\label{aa-w}
\\[0.2cm]
a^*&=&q^{+\frac{1}{2}}\,(U-U^{-1})^{-1}\, (U\,V^{-1}-U^{-1}\,V),\nonumber
\end{eqnarray}
where the elements $U$ and $V$ generate the Weyl algebra,
\be
U\,V=q\,V\,U,\qquad U=e^{i\alpha},\qquad
V=e^{i\beta},\qquad q=e^{-\xi\hbar}\ .\label{weyl1}
\ee
The operators \eqref{aa-w} obey the commutation relations of the
$q$-oscillator algebra,
\be\label{q-osc1}
\mathsf{Osc}_{\,q}:\qquad
\left\{
\renewcommand\arraystretch{2.0}
\begin{array}{l}
q\,a^* \, a- q^{-1}\, a \,a^*=q-q^{-1},
\qquad k\,a^*=q\,a^*\,k,\qquad k\,a=q^{-1}\,a\,k\;,\\
k^2=q\,(1-a^*\, a)=q^{-1}\,(1-a\, a^*)\ .
\end{array}\right.
\renewcommand\arraystretch{1.0}
\ee
where the element $k$ is assumed to be invertible.
This algebra is, obviously,
a quantum counterpart of the
Poisson algebra (\ref{o-poisson}).
In the previous Section we have already mentioned the result of
\cite{BS05} that
\begin{itemize}
\item[(i)] the map \eqref{themap} is
an automorphism of the tensor cube of the Poisson algebra
\eqref{o-poisson} (remind that the relation \eqref{aak-rel} should be
taken into account in \eqref{themap}).
\end{itemize}
In the same paper \cite{BS05} it was also shown that
\begin{itemize}
\item[(ii)]
there exists a quantum version of the map \eqref{themap}, which
acts as an automorphism of the tensor cube of
the $q$-oscillator algebra \eqref{q-osc1}.
The formulae \eqref{themap} for the quantum map stay exactly the same, but
the relation
\eqref{aak-rel} should be replaced by either of the two relations on
the second line of \eqref{q-osc1}, for instance,
$k^2=q\,(1-a^* a)$. In particular, \eqref{aak2} should be replaced with
\be
(k_2')^2=q\,(1-{a_2^*}^{\prime} {a_2}')\ .\label{aak2q}
\ee
\end{itemize}

For irreducible representations of the $q$-oscillator algebra
\eqref{q-osc1} the formulae \eqref{themap} and \eqref{aak2q}
uniquely determine the
$\R_{123}$ as
an internal automorphism,
\begin{equation}\label{rl-def}
\R_{123}\big( F \big)\;=\; R_{123}^{} \;F\; R_{123}^{-1},\qquad
 F\in \mathsf{Osc}_q\otimes\mathsf{Osc}_q\otimes\mathsf{Osc}_q\ .
\end{equation}
It follows then from \eqref{fte} that the linear operator $R$
 satisfies the quantum tetrahedron equation
\begin{equation}\label{TE}
R_{123}\;R_{145}\;R_{246}\;R_{356}\;=\;
R_{356}\;R_{246}\;R_{145}\;R_{123}\;,
\end{equation}
where the operators $R_{123}$, $R_{145}$, $R_{246}$ and $R_{356}$ act
as \eqref{rl-def} in the three factors of the tensor product
of the six algebras $q$-oscillator algebras
(indicated by their subscripts) and
act as the unit operator in the remaining three factors.

\subsection{Fock space representation model}

In this subsection we set $\xi=1$ in \eqref{ang-alg}. Assuming
$\hbar>0$ we have $q=e^{-\hbar}<1$.
Define the Fock space representation of a single $q$-oscillator
algebra \eqref{q-osc1},
\begin{equation}
{\mathsf F}_q: \qquad
a\,|0\rangle =0\;,\quad |n\rangle = \frac{(a^*)^n}{\sqrt{(q^2;q^2)_n}}\,
|0\rangle\;,\quad k\,|n\rangle = q^{n+1/2}\,|n\rangle \;,\quad
n\geq 0
\end{equation}
Then using \eqref{themap}, \eqref{q-osc1}, \eqref{aak2q}
and \eqref{rl-def} one can show that the
the matrix elements of $R$ are given by
\begin{equation}\label{R-fock}
\renewcommand\arraystretch{2.2}
\begin{array}{ll}
\ds &\langle n_1,n_2,n_3|R\,|n'_1,n'_2,n_3\rangle =
\ds \delta_{n_1^{}+n_2^{},n_1'+n_2'}
\delta_{n_2^{}+n_3^{},n_2'+n_3'}\;
\sqrt{\frac{(q^2;q^2)_{n_1'}\,(q^2;q^2)_{n_2'}\,(q^2;q^2)_{n_3'}}
{(q^2;q^2)_{n_1}\,(q^2;q^2)_{n_2}\,(q^2;q^2)_{n_3}}}\\[.3cm]
&\ds\times \,\frac{(-1)^{n_2}\,q^{(n_1'-n_2^{})(n_3'-n_2^{})}}
{(q^2;q^2)_{n_2'}}\;
\frac{(q^{2(1-n_2'+n_3)};q^2)_\infty}{(q^{2(1+n_3)};q^2)_\infty}\,
\,{}_2\phi_1(q^{-2n_2'},q^{2(1+n_3')},q^{2(1-n_2'+n_3)};q^2,q^{2(1+n_1)})\ ,
\end{array}
\renewcommand\arraystretch{1.0}
\end{equation}
where
\begin{equation}
(x;q^2)_n=(1-x)(1-q^2x)\cdots
(1-q^{2(n-1)}x)\ ,
\end{equation}
and
\begin{equation}\label{q-gauss}
\phantom{f}_2\phi_1(a,b,c;q^2,z)\;=\;
\sum_{n=0}^{\infty}
\frac{(a;q^2)_n(b;q^2)_n}{(q^2;q^2)_n(c;q^2)_n} z^n
\end{equation}
is the $q$-deformed Gauss hypergeometric series.
Eq.\eqref{rl-def} together with \eqref{themap} and \eqref{aak2q} lead
to recurrence relations for the matrix elements of $R$.
Such relations along with
expressions for matrix elements of $R$ for small values of the occupation
numbers $n_i$, were obtained in \cite{BS05} (see Eqs.(30-32) therein). We
have now solved those recurrence relations and obtained the
explicit formula \eqref{R-fock}, given above\footnote{%
  Eq.\eqref{R-fock} follows from Eqs.(29,30) of \cite{BS05} where
$P_{\beta'}(q^{2\alpha},q^{2\beta},q^{2\gamma})$ is substituted with
$$
P_{\beta'}(q^{2\alpha},q^{2\beta},q^{2\gamma})=
\frac{(q^{2(1-\beta'+\gamma)};q^2)_\infty}
     {(q^{2(1+\gamma)};q^2)_\infty} \,
{}_2\phi_1(q^{-2\beta'},q^{2(1-\beta'+\beta+\gamma)},
q^{2(1-\beta'+\gamma)};q^2,q^{2(1+\alpha)}).
$$}%
.
This 3D $R$-matrix satisfies the constant tetrahedron equation
(\ref{TE}),
\begin{equation}\label{te-fock}
\sum_{n_j'=0}^\infty R_{n_1^{},n_2^{},n_3^{}}^{n_1',n_2',n_3'}\,
R_{n_1',n_4^{},n_5^{}}^{n_1'',n_4',n_5'}\,
R_{n_2',n_4',n_6^{}}^{n_2'',n_4'',n_6'}\,
R_{n_3',n_5',n_6'}^{n_3'',n_5'',n_6''} = \sum_{n_j'=0}^\infty
R_{n_3^{},n_5^{},n_6^{}}^{n_3',n_5',n_6'}\,
R_{n_2^{},n_4^{},n_6'}^{n_2',n_4',n_6''}\,
R_{n_1^{},n_4',n_5'}^{n_1',n_4'',n_5''}\,
R_{n_1',n_2',n_3'}^{n_1'',n_2'',n_3''}
\end{equation}
where the sum is taken over six indices
$n'_1,n'_2,n'_3,n'_4,n'_5,n'_6$ and
\begin{equation}
R_{n_1,n_2,n_3}^{n_1',n_2',n_3'}=\langle n_1,n_2,n_3|R\,|n_1',n_2',n_3'\rangle
\ .
\end{equation}
Note that Eq.\eqref{te-fock} does not contain any spectral parameters.

\begin{figure}[ht]
\vspace{1cm}
\begin{center}
\setlength{\unitlength}{0.35mm}
\begin{picture}(140,140)
\put(20,20)
 {\begin{picture}(100,100)
 \path(50,0)(50,100)\path(45,10)(50,0)(55,10)
 \path(0,50)(100,50)\path(10,45)(0,50)(10,55)
 \path(10,30)(90,70)\path(80,70)(90,70)(80,60)
 \put(45,-15){\scriptsize $n_1'$}\put(45,110){\scriptsize $n_1$}
 \put(-15,48){\scriptsize $n_2'$}\put(110,48){\scriptsize $n_2$}
 \put(-5,20){\scriptsize $n_3$}\put(95,75){\scriptsize $n_3'$}
 \put(50,50){\circle*{5}}
 \end{picture}}
\end{picture}
\caption{Graphical visualization of three-dimensional $R$-matrix \eqref{R-fock}
and its matrix elements. The orientation of the axes is consistent with
Fig.~\ref{fig-cube2}} \label{fig-3DR2}
\end{center}
\end{figure}
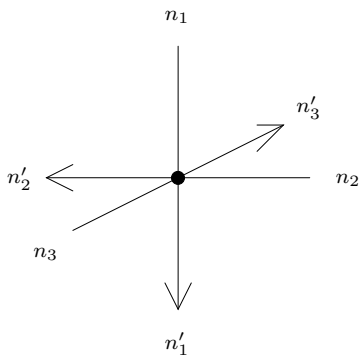

Now define a model of lattice field theory.
On each edge of the cubic lattice
place a fluctuating spin variable $n$, taking an infinite
number of integer values $n=0,1,2,\ldots\infty$. To each vertex of the lattice
assign a local weight factor $\langle
n_1,n_2,n_3|R\,|n_1',n_2',n_3'\rangle$, given by \eqref{R-fock},
 depending on six spin variables
placed on the edges surrounding the vertex, arranged as in
Fig.\ref{fig-3DR2}.
The partition function
\be
Z_{\mathsf F}=\sum_{\mathrm{(spins)}} \ \prod_{\mathrm{(vertices)}}
\langle n_1,n_2,n_3|R\,|n_1',n_2',n_3'\rangle\ ,\label{pf-f}
\ee
is defined as the sum over all spin configurations of the whole lattice where
each configuration is counted with the weight equal to the product of
the vertex weights over all lattice vertices. For definiteness we
assume fixed boundary condition. The subscript ``${\mathsf F}$''  stands
for the ``Fock representation model''.

In the quasi-classical limit
\be
q=e^{-\hslash}\to1,\qquad \hslash\to 0,\qquad n_i\hslash = -\log k_i =
\textrm{finite}, \label{klim-def}
\ee
with large occupation numbers $n_i$, $n'_i$, such that $k_i$ and $k_i'$
are kept finite, one obtains
\begin{equation}
\lim_{\hslash\to 0}\, \langle n_1,n_2,n_3|R\,|n_1',n_2',n_3'\rangle
= \exp\Big( - \frac{\overline{{\cal L}}_{{\mathsf
F}}(k,k')}{\hbar}\Big)\;,\label{fc-lim}
\end{equation}
where the arguments of $\overline{\cal L}_{\mathsf F}(k,k')$ are related by
$k_1^{}k_2^{}=k_1'k_2'$ and $k_2^{}k_3^{}=k_2'k_3'$.
The function $\overline{\cal L}_{\mathsf F}(k,k')$ coincides with the classical
Lagrangian \eqref{S-1} provided the variables
$k,k'\/$  in \eqref{klim-def} are
identified with those in \eqref{kv}. The sum \eqref{pf-f} in this
limit can be replaced by an integral and estimated 
by the saddle point method, 
\be\label{q-limit}
\log Z_{\mathsf F} = - \frac {1}{\hbar}\,{\cal
  S}^{(cl)}+O(\hbar^0),\qquad \hbar\to0,
\ee
where ${\cal S}^{(cl)}$ is the classical action \eqref{actk}
evaluated at its stationary point.

\subsection{Modular double model}\label{modular-sec}

In this subsection we set in \eqref{ang-alg}
\be
\xi=-i,\qquad \hbar=\pi\,b^2,
\ee
where $b$ is a free parameter, $\mathrm{Re}\, b\not=0$. Here it will
be more
convenient to work with a slightly modified version\footnote{%
Note that the map \eqref{themap} with $\varepsilon =\pm1$ is a particular
case a three-parameter map considered in \cite{BS05}.}
of the map \eqref{themap}, with the value $\varepsilon =-1$.
It is worth noting that this modification does not affect the bulk
properties of the classical system \eqref{actv} and leads only to
boundary effects. In particular the equations of motion
\eqref{eight-bra} remain unchanged.

Consider a non-compact representation \cite{Schmud94} of the $q$-oscillator
algebra \eqref{q-osc1} in the space of functions $f(\sigma)\in {\mathbb
  L}^2({\mathbb R})$ on the real line admitting an analytical
continuation into an appropriate horizontal
strip, containing the real axis in the
complex $\sigma$-plane (see
\cite{Schmud94} and \cite{BMS08} for further details).
Such representation essentially reduces to that of the Weyl algebra
\be
{\mathsf W}_q:\qquad\qquad k\,w =q\, w\, k,\qquad q=e^{i\pi b^2},\label{weyl2}
\ee
realized as multiplication and shift operators
\be
k\,|\sigma\rangle = \ii
e^{\pi\sigma b}\, |\sigma\rangle\;,\qquad w\,
|\sigma\rangle = |\sigma-\ii b\rangle. \label{w-rep}
\ee
The generators $a,a^*$ in \eqref{q-osc1} are expressed as
\be
a=(1-q\,k^2)^{1/2}\,w^{-1},\qquad a^* =(1-q^{-1}k^2)^{1/2}\,w.
\ee
As explained in \cite{Faddeev:1995} the representation \eqref{w-rep}
is not, in general, irreducible. Therefore, the relation
\eqref{rl-def} alone does not unambiguously define 
the linear operator $R_{123}$ in this case. 
Following the idea of \cite{Faddeev:1995} consider the modular 
dual of the algebra \eqref{weyl2}, 
\be
{\mathsf W}_{\tilde{q}}:\qquad\qquad \tilde{k}\,\tilde{w} =\tilde{q}\,
\tilde{w}\, \tilde{k},\qquad \tilde{q}=e^{-i\pi b^{-2}},\label{weyl3}
\ee
acting in the same representation space
\be
\tilde{k}\,|\sigma\rangle = -\ii
e^{\pi\sigma b^{-1}}\, |\sigma\rangle\;,\qquad \tilde{w}\,
|\sigma\rangle = |\sigma+\ii b^{-1}\rangle.
\ee
We found that if the 
relation \eqref{rl-def} is complemented by its modular dual
\begin{equation}\label{rl-def1}
\tilde{\R}_{123}\big( \tilde{F} \big)
\;=\; R_{123}^{} \;\tilde{F}\; R_{123}^{-1},\qquad
 \tilde{F}\in
 \mathsf{Osc}_{\tilde{q}}
\otimes\mathsf{Osc}_{\tilde{q}}\otimes\mathsf{Osc}_{\tilde{q}}\ ,
\end{equation}
then the pair of relations \eqref{rl-def} and \eqref{rl-def1}
determine the operator $R_{123}$ uniquely\footnote{It is worth
  mentioning similar phenomena in
  the construction of the $R$-matrix \cite{Faddeev:1999} for the
  modular double of the quantum group $U_q(sl_2)$ and the
  representation theory of $U_q(sl_2,{\mathbb R})$ \cite{PT99}.}.
The dual $q$-oscillator
algebra $\mathsf{Osc}_{\tilde{q}}$
is realized through the dual Weyl pair \eqref{weyl3} and the relations
\be
\tilde{a}=(1-\tilde{q}\,\tilde{k}^2)^{1/2}\,\tilde{w}^{-1},\;\;
\tilde{a}^* =(1-\tilde{q}^{-1}\tilde{k}^2)^{1/2}\, \tilde{w}.
\end{equation}
The dual version of the map $\tilde{\R}_{123}$ is defined by the same
formulae \eqref{themap}, where quantities
$k_j,a_j,a^*_j$, \ $j=1,2,3$ are replaced by their ``tilded''
counterparts $\tilde{k}_j,\tilde{a}_j,\tilde{a}^*_j$. The value of $q$
does not, actually, enter the map \eqref{themap}, but needs to be taken
into account in the relations between the generators of the
$q$-oscillator algebra.
Thus, the linear operator $R_{123}$ in this case simultaneously provides
the two maps ${\R}_{123}$ and $\tilde{\R}_{123}$ (with
$\varepsilon=-1$). The explicit form of this operator is given below.

Denote
\be
\eta=\frac{b+b^{-1}}{2},
\ee
and define a special function
\begin{equation}\label{spec-f}
\Bigpsi{c_1,c_2}{c_3,c_4}{c_0}\;=\; \int_{\mathbb{R}} \, dz\,
e^{2\pi\ii z (-c_0-\ii\eta)}\,
\frac{\varphi(z+\frac{c_1+\ii\eta}{2})\varphi(z+\frac{c_2+\ii\eta}{2})}
{\varphi(z+\frac{c_3-\ii\eta}{2})\varphi(z+\frac{c_4-\ii\eta}{2})}\;,
\end{equation}
where $\varphi$ is the non-compact quantum dilogarithm
\cite{Faddeev:1994}
\begin{equation}
\varphi(z)=\exp\left(\ds \frac{1}{4}\int_{\mathbb{R}+\ii 0}
\frac{e^{-2\ii zx}}{\textrm{sinh}(xb)\textrm{sinh}(x/b)}\
\frac{dx}{x}\right)\;.
\end{equation}
The values of $c_1,c_2,c_3,c_4$ are assumed to be such that poles of
numerator in the integrand of \eqref{spec-f} lie above the real axis,
while the zeroes of the denominator lie below the real axis. For other
values of $c_j$ the integral \eqref{spec-f} is defined by an analytic
continuation.
For $\mathrm{Im}\,b^2>0$ the integral
$\phantom{|}_2 \kern -.05em \Psi_2$ can be
evaluated by closing the integration contour in the
upper half plane. The result reads
\be
\renewcommand\arraystretch{2.0}
\begin{array}{l}
\ds\Bigpsi{c_1,c_2}{c_3,c_4}{c_0}\;=\;\ds
e^{i\pi(c_0+i\eta)(c_1-i\eta)-i\pi(4\eta^2+1)/12}\
\frac{\ds
\varphi\Big(\frac{c_2-c_1}{2}+i\eta\Big)}
{\ds\varphi\Big(\frac{c_3-c_1}{2}\Big)\,
\varphi\Big(\frac{c_4-c_1}{2}\Big)}
\\[.3cm]
\phantom{\Psi\Psi\Psi\Psi}\times{}_2\phi_1(-\tilde q\,e^{\pi(c_1-c_3)/b},
-\tilde q\, e^{\pi(c_1-c_4)/b},
\tilde q^2\,e^{\pi(c_1-c_2)/b};\tilde q^2,
-\tilde q\,e^{\pi(c_3+c_4-c_1-c_2+2c_0)/b})\\[.3cm]
\phantom{\Psi\Psi\Psi\Psi}
\times{\ds
{}_2\phi_1(-q\,e^{\pi (c_3-c_1)b},-q\,e^{\pi (c_4-c_1)b},
q^2\,e^{\pi (c_2-c_1)b};q^2,-q\,e^{2\pi b c_0})\ +\
(c_1\leftrightarrow c_2),}
\end{array}\label{2terms}
\renewcommand\arraystretch{1.0}
\ee
where ${}_2\phi_1$ is defined by \eqref{q-gauss}.
This formula is very convenient
for numerical calculations.

The kernel of $R$-matrix serving the pair of the maps \eqref{rl-def}
and \eqref{rl-def1} is given by
\begin{equation}\label{R-modular}
\renewcommand\arraystretch{3.0}
\begin{array}{rcl}
\langle \sigma |R\,|\sigma'\rangle &=& \ds
\delta_{\sigma_1^{}+\sigma_2^{},\sigma_1'
+\sigma_2'}\delta_{\sigma_2^{}+\sigma_3^{},\sigma_2'+\sigma_3'}
e^{\ii \pi
(\sigma_1'\sigma_3'+\ii\eta(\sigma_1'+\sigma_3'-\sigma_2^{}))}\\
&&\times\ds
\sqrt{\frac{\varphi(\sigma_1)\varphi(\sigma_2)\varphi(\sigma_3)}
{\varphi(\sigma_1')\varphi(\sigma_2')\varphi(\sigma_3')}}
\Bigpsi{\sigma_1-\sigma_3,-\sigma_1+\sigma_3}
{\sigma_1+\sigma_3,-\sigma_1'-\sigma_3'}{\sigma_2}
\end{array}
\renewcommand\arraystretch{1.0}
\end{equation}
It satisfies the constant
tetrahedron equation (\ref{TE}),
\begin{equation}
\begin{array}{l}
\ds \int_{\mathbb{R}}d\sigma_1'...d\sigma_6'
\;R_{\sigma_1^{\phantom{\prime}},\sigma_2^{},
\sigma_3^{}}^{\sigma_1',\sigma_2',\sigma_3'}
\;R_{\sigma_1',\sigma_4^{},\sigma_5^{}}^{\sigma_1'',\sigma_4',\sigma_5'}
\;R_{\sigma_2',\sigma_4',\sigma_6^{}}^{\sigma_2'',\sigma_4'',\sigma_6'}
\;R_{\sigma_3',\sigma_5',\sigma_6'}^{\sigma_3'',\sigma_5'',\sigma_6''}=\\
\\[.2cm]
\ds \phantom{
\int_{\mathbb{R}}d\sigma_1'...d\sigma_6'
\;R_{\sigma_1^{\phantom{\prime}},\sigma_2^{},
\sigma_3^{}}^{\sigma_1',\sigma_2',\sigma_3'}
}
= \int_{\mathbb{R}}d\sigma_1'...d\sigma_6'
\;R_{\sigma_3^{\phantom{\prime}},\sigma_5^{},
\sigma_6^{}}^{\sigma_3',\sigma_5',\sigma_6'}
\;R_{\sigma_2^{},\sigma_4^{},\sigma_6'}^{\sigma_2',\sigma_4',\sigma_6''}
\;R_{\sigma_1^{},\sigma_4',\sigma_5'}^{\sigma_1',\sigma_4'',\sigma_5''}
\;R_{\sigma_1',\sigma_2',\sigma_3'}^{\sigma_1'',\sigma_2'',\sigma_3''}
\end{array}
\end{equation}
where
\begin{equation}
\langle
\sigma_1,\sigma_2,\sigma_3|R\,|\sigma_1',\sigma_2',\sigma_3'\rangle
\;=\;
R_{\sigma_1,\sigma_2,\sigma_3}^{\sigma_1',\sigma_2',\sigma_3'}
\end{equation}
and the integrals are taken along the real line from $-\infty$ to $+\infty$.

Similarly to \eqref{pf-f} define a ``modular double model''
\be
Z_{\mathsf M}=\int \cdots \int \ \prod_{\mathrm{(vertices)}}
\langle \sigma|R\,|\sigma'\rangle\
\prod_{\mathrm{(edges)}} d\sigma\ ,\label{pf-m}
\ee
where the edge spins $\sigma$ now take continuous values on the real
line. We assume fixed boundary conditions.
Note that due to the presence of two delta-functions in \eqref{R-modular}
the edge spins are constrained by two
relations
\begin{equation}
\sigma_1^{}+\sigma_3^{}=\sigma_1'+\sigma_3',\qquad
\sigma_2^{}+\sigma_3^{}=\sigma_2'+\sigma_3',\label{sig-cons}
\end{equation}
at each vertex of the lattice.

Consider the quasi-classical limit of the $R$-matrix \eqref{R-modular}
\be
b\to0,\qquad |\sigma_j|,|\sigma'_j|\to\infty,\label{qcl}
\ee
such that the variables
\be
k_j=ie^{\pi b \sigma_j},\qquad k'_j=ie^{\pi b \sigma'_j},\qquad
j=1,2,3,
\label{k-var1}
\ee
are kept finite. Anticipating a connection with the classical formula
\eqref{S-1} let us identify the variables \eqref{k-var1} with those in
\eqref{kv}. Further, for real $b$ the squares $k_j^2$ and ${k'_j}^2$
are negative (notice the factor $i$ in front of the exponents in
\eqref{k-var1}). To accommodate this extra minus sign we will use
a slightly modified form of \eqref{v2E}, namely,
\begin{equation}
\begin{array}{llll}
\ds e^{2\ii\omega_2}=\frac{v_2'}{v_1'v_3'}\;, & \ds
e^{2\ii\omega_1}=\frac{v_2}{v_1v_3'}\;, & \ds
e^{2\ii\omega_0}=\frac{v_2'}{v_1v_3}\;, &
\ds e^{2\ii\omega_3}=\frac{v_2}{v_1'v_3}\;,\\
&&&\\
\ds e^{-2\ii\omega_2}=\frac{v_2}{v_1v_3}\;, & \ds
e^{-2\ii\omega_1}=\frac{v_2'}{v_1'v_3}\;, & \ds
e^{-2\ii\omega_0}=\frac{v_2}{v_1'v_3'}\;, & \ds
e^{-2\ii\omega_3}=\frac{v_2'}{v_1v_3'}\;.
\end{array}\label{v2om}
\end{equation}
where the new variables $\omega_j$ and $\omega_j'$ obey the relations
\be
\omega'_k=\frac{\pi}{2}+
\omega_k-(\omega_0+\omega_1+\omega_2+\omega_3)/2\label{o-rel},
\qquad k=0,1,2,3\ .
\ee
Note also a useful identity
\begin{equation}
\omega_i+\omega_j+\omega_k'+\omega_l'=\pi,\qquad
(i,j,k,l)=\mathrm{perm}(0,1,2,3),
\end{equation}
where $(i,j,k,l)$ is any permutation of $(0,1,2,3)$. If
the $\omega$-variables lie in the domain
\be
0<\omega_k<\pi,\qquad 0<\omega_k'<\pi, \qquad k=0,1,2,3,
\ee
then in the quasi-classical limit \eqref{qcl} the
integral \eqref{spec-f} entering the formula
\eqref{R-modular} can be estimated by the saddle point method
\begin{equation}
\langle\sigma|R\,|\sigma'\rangle =
\exp\left\{-\frac{\overline{\cal L}_{\mathsf{M}}(k,k')}{\pi
b^2}+O(1)\right\}\;,\qquad {b\to 0}\ .\label{mod-sad}
\end{equation}
In practice, it is easier to calculate logarithmic derivatives of
\eqref{R-modular} with respect variables $k_j$ and $k'_j$,
rather than the expression \eqref{R-modular} itself. In this way
one obtains
\begin{equation}
d\overline{\mathcal{L}}_\mathsf{M}(k,k')\;=\;\frac{1}{2}\sum_{j=0}^3
\left(\omega_jd\log\sin\omega_j +
\omega_j'd\log\sin\omega_j'\right)
-\frac{\pi}{2}\left(d\log\sin\omega_2+d\log\sin\omega_2'\right)\ .
\end{equation}
It follows then
\begin{equation}
\overline{\mathcal{L}}_{\mathsf{M}}(k,k') \;=\;
\mathcal{L}_{\mathsf{M}}(v,v') + \textrm{Legendre terms}\ ,
\end{equation}
where the non-trivial part
\begin{equation}
\mathcal{L}_{\mathsf{M}}(v,v')\;=\;\frac{1}{2}\sum_{j=0}^3
\left(\Lb(\omega_j)+\Lb(\omega_j')\right)\ ,\label{lm}
\end{equation}
is expressed in terms of the Lobachevsky function \cite{milnor},
\begin{equation}
\Lb(\omega)\;=\; -\int_0^\omega \log 2\sin x \;dx\;,\quad
0<\omega<\pi\;,
\end{equation}
while the Legendre terms read
\begin{equation}
\textrm{Legendre terms} = \frac{1}{2}\sum_{j=0}^3
\left(\omega_j\log 2\sin\omega_j +
\omega_j'\log2\sin\omega_j'\right) -
\frac{\pi}{2}\left(\log2\sin\omega_2+\log 2\sin\omega_2'\right)\ .
\end{equation}
Note, that in terms of the variables $k,k',v,v'$ (connected with $\omega$'s
by \eqref{kv} and  \eqref{v2om}) the differential of \eqref{lm} reads
\begin{equation}
d\mathcal{L}_{\mathsf{M}}(v,v')=\frac{1}{2\ii} \sum_{j=1}^3 \left(
\log \frac{k_j}{\ii}d\log v_j - \log\frac{k_j'}{\ii} d\log
v_j'\right)\ ,\label{diff1}
\end{equation}
while the Legendre terms take the form
\begin{equation}
\textrm{Legendre terms} = -\frac{1}{2\ii}\sum_{j=1}^3 \left( \log
\frac{k_j}{\ii} \log v_j - \log\frac{k_j'}{\ii}  \log v_j'\right)\
.\label{leg1}
\end{equation}
The formulae \eqref{diff1} and \eqref{leg1} are in a complete agreement
with \eqref{dS} and \eqref{S-1}, except the overall normalization and the
replacement $k_j\to k_j/\ii$ and $k_j'\to k_j'/\ii$ which is a gauge
transformation resulting from a different of choice of the canonical
variables. The total action
\be
{\mathcal   S}^{(cl)}_M(\{k\})=\sum_{m\in{\mathbb Z}^3}
{\mathcal L}_{\mathsf M}(k(m),k'(m))\label{actm-k}
\ee
coincides (to within an overall factor $i^{-1}$ and boundary terms) with
an analytic continuation of \eqref{actk} into a regime where the
variables $k_j^2$ are negative and the variables $v_j$ are unimodular.
Substituting \eqref{mod-sad} into \eqref{pf-m} one obtains
\be
\log Z_{\mathsf M}=-\frac{1}{\pi b^2}\, {\mathcal
  S}^{(cl)}_M+O(1),\qquad b\to0
\ee
where ${\mathcal   S}^{(cl)}_M$ is the classical action \eqref{actm-k}
evaluated at its stationary point. It is obvious from \eqref{lm} that
this expression is real.

Before concluding this section let us make a remark about the spatial
symmetry properties of the modular model.
It is convenient to introduce another kernel,
\begin{equation}\label{R-symmetric}
\langle \sigma |\widetilde{R}|\sigma'\rangle = \langle\sigma
|R\,|\sigma'\rangle \; e^{\pi\eta
(\sigma_1^{}-\sigma_2^{}+\sigma_3^{}+\sigma_1'-\sigma_2'+\sigma_3')/2}
\end{equation}
differing from (\ref{R-modular}) by a phase factor which introduces
particular
``external fields'' but does not destroy the integrability of the model.
The new kernel $\widetilde{R}$ obeys simple relations
\begin{equation}
\widetilde{R}_{\sigma_1^{\phantom{\prime}},
\sigma_2^{},\sigma_3^{}}^{\sigma_1',\sigma_2',\sigma_3'}\;=\;
\widetilde{R}_{\sigma_3^{\phantom{\prime}},
\sigma_2^{},\sigma_1^{}}^{\sigma_2',\sigma_2',\sigma_1'}\;=\;
\widetilde{R}_{-\sigma_2',\sigma_1^{},
-\sigma_3^{}}^{-\sigma_2^{},\sigma_1',-\sigma_3'}\;.
\end{equation}
which generate the whole cube symmetry group.

\subsection{The ``interaction-round-a-cube'' formulation of the
  modular model}\label{IRC}

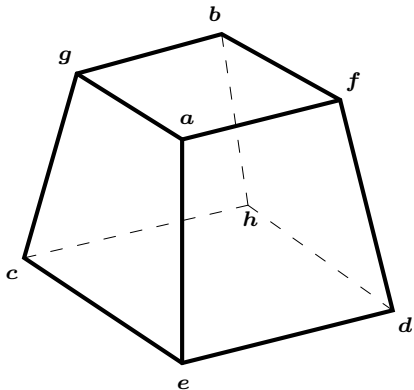
\begin{figure}[ht]
\centering
\setlength{\unitlength}{0.35mm}
\begin{picture}(150,150)
\put(0,10){
\begin{picture}(150,125)
\allinethickness{.6mm}
 \path(0,40)(60,0)(140,20)(120,100)(75,125)(20,110)(0,40)
 \path(60,85)(20,110)
 \path(60,85)(120,100)
 \path(60,85)(60,0)
\thinlines
 \dashline{5}(85,60)(0,40)
 \dashline{5}(85,60)(75,125)
 \dashline{5}(85,60)(140,20)
 \put(59,91){\scriptsize $\boldsymbol{a}$}
 \put(58,-10){\scriptsize $\boldsymbol{e}$}
 \put(-7,32){\scriptsize $\boldsymbol{c}$}
 \put(142,12){\scriptsize $\boldsymbol{d}$}
 \put(13,115){\scriptsize $\boldsymbol{g}$}
 \put(122,105){\scriptsize $\boldsymbol{f}$}
 \put(70,130){\scriptsize $\boldsymbol{b}$}
 \put(83,52){\scriptsize $\boldsymbol{h}$}
\end{picture}}\end{picture}
\caption{The arrangement of corner spins
around a cube.}\label{fig-cubeweight}
\end{figure}

An inconvenient feature of the modular representation model formulated
above is that the edge spins are constrained by the relations
\eqref{sig-cons}. Here we re-formulate this model
in terms of unconstrained {\em corner\/}\  spins, which also
take continuous values on the real line.
Fig.~\ref{fig-cubeweight} shows an elementary cube of
the lattice with the corner spins ``$a,b,c,d,e,f,g,h$'' arranged in the
same way as in \cite{Baxter:1986phd}.
The corresponding Boltzmann weight reads

\begin{eqnarray}
&
W(a|e,f,g|b,c,d|h;
\T_1,\T_2,\T_3;\a_1,\a_2,\a_3) \;=\;
\exp{\Bigl[{\ds\sum_{i=1}^3\a_i(\sigma_i+\sigma_i')
\eta(-1)^{\delta_{i,3}}}\Bigr]}\times&\nonumber\\
& \ds\times e^{-\ii \pi
(\sigma_1\sigma_2+\ii\eta(\sigma_3'-\sigma_1-\sigma_2))}
\sqrt{\frac{\varphi(\sigma_1)\varphi(\sigma_2)\varphi(\sigma_3)}
{\varphi(\sigma_1')\varphi(\sigma_2')\varphi(\sigma_3')}}
\Bigpsi{\sigma_1'+\sigma_2',-\sigma_1-\sigma_2}{\sigma_1-\sigma_2,-\sigma_1+\sigma_2}{-\sigma_3'}
\label{Wweight}
\end{eqnarray}
where  $\sigma_i$, $\sigma_i'$ are defined as
\begin{equation}
\begin{array}{lll}
\sigma_1=g+f-a-b-\T_1\;,& \sigma_2=e+g-a-c-\T_2\;, &
\sigma_3=a+d-e-f+\T_3\;,\\&&\\ \sigma_1'=c+d-e-h-\T_1\;,&
\sigma_2'=b+d-f-h-\T_2\;, & \sigma_3'=g+h-b-c+\T_3
\end{array}
\end{equation}
and satisfy the constrains \eqref{sig-cons}.
The parameters $\T_1,\T_2,\T_3$ are analogs of the spectral parameters in the
Zamolodchikov model \cite{Zamolodchikov:1981kf} and its
generalization for an arbitrary number $N\ge2$
of spin states \cite{Bazhanov:1992jq}.
They are related by $\T_j=\log[\tan(\theta_j/2)]$, where $\theta_j$, \
$j=1,2,3$,\  are the dihedral angles in the Zamolodchikov model (they are
usually considered as angles of a spherical triangle).
The parameters $\a_i$ are similar to the linear
angles in Zamolodchikov model (the side lengths of the above spherical
triangle).
However, unlike the Zamolodchikov model, the two sets of parameters
$\T_i$ and $\a_i$ in our case are totally independent.
The parameters $\a_i$ enter only
the exponential prefactor in (\ref{Wweight}); they play the role of
the external fields in the model.

Consider spatial symmetry properties of the weight function
\eqref{Wweight}.
The group of cube is generated by
two elements: $\tau$ (the reflection with respect to the diagonal plane
$(a,d,h,g)$ in Fig.~\ref{fig-cubeweight})
and $\rho$ (the $90^\circ$ rotation around the vertical axis).
They act on the spin variables and
parameters $\T_i$, $\a_i$ as follows
\be\label{tau1}
\renewcommand\arraystretch{2.0}
\begin{array}{l}
\tau: \ (a|efg|bcd|h;\T_1,\T_2,\T_3,\a_1,\a_2,\a_3)\to
(a|feg|cbd|h;\T_2,\T_1,\T_3;\a_2,\a_1,\a_3),\\
\rho: \ (a|efg|bcd|h;\T_1,\T_2,\T_3,\a_1,\a_2,\a_3)\to
(g|cab|fhe|d;-\T_1,\T_3,-\T_2;\pi-\a_1,\a_3,\pi-\a_2).
\end{array}
\renewcommand\arraystretch{1.0}
\ee
On can show show that the weight function (\ref{Wweight})
is invariant under these substitutions and, thus, possesses
the full cube symmetry group.
\begin{figure}[ht]
\centering
\addtolength{\unitlength}{-0.5\unitlength}
\begin{picture}(700,440)
\put(50,100){\begin{picture}(300,300)
\thicklines
\multiput(141,0)(60,100){2}{\line(2,1){80}}
\multiput(140,0)(60,100){2}{\line(2,1){80}}
\multiput(139,0)(60,100){2}{\line(2,1){80}}
\multiput(140,1)(80,40){2}{\line(3,5){60}}
\multiput(140,0)(80,40){2}{\line(3,5){60}}
\multiput(140,-1)(80,40){2}{\line(3,5){60}}
\multiput(0,141)(60,-40){2}{\line(3,5){60}}
\multiput(0,140)(60,-40){2}{\line(3,5){60}}
\multiput(0,139)(60,-40){2}{\line(3,5){60}}
\multiput(1,140)(60,100){2}{\line(3,-2){60}}
\multiput(0,140)(60,100){2}{\line(3,-2){60}}
\multiput(-1,140)(60,100){2}{\line(3,-2){60}}
\multiput(0,141)(60,-40){2}{\line(4,-5){80}}
\multiput(0,140)(60,-40){2}{\line(4,-5){80}}
\multiput(0,139)(60,-40){2}{\line(4,-5){80}}
\multiput(200,101)(80,40){2}{\line(-4,5){80}}
\multiput(200,100)(80,40){2}{\line(-4,5){80}}
\multiput(200, 99)(80,40){2}{\line(-4,5){80}}
\multiput(61,240)(60,-40){2}{\line(2,1){80}}
\multiput(60,240)(60,-40){2}{\line(2,1){80}}
\multiput(59,240)(60,-40){2}{\line(2,1){80}}
\multiput(141,0)(60,240){2}{\line(-3,2){60}}
\multiput(140,0)(60,240){2}{\line(-3,2){60}}
\multiput(139,0)(60,240){2}{\line(-3,2){60}}
\multiput(160,80)(30,-20){2}{\line(3,-2){20}}
\multiput(220,180)(30,-20){2}{\line(3,-2){20}}
\multiput(80,40)(30,15){3}{\line(2,1){20}}
\multiput(0,140)(30,15){3}{\line(2,1){20}}
\multiput(160,80)(-20,25){4}{\line(-4,5){16}}
\multiput(220,180)(-20,25){4}{\line(-4,5){16}}
\multiput(220,180)(-16,-26.667){4}{\line(-3,-5){12}}
\multiput(140,280)(-16,-26.667){4}{\line(-3,-5){12}}
\multiput(140,140)(30,-20){2}{\line(3,-2){20}}
\multiput(140,140)(30,15){3}{\line(2,1){20}}
\multiput(140,140)(-20,25){4}{\line(-4,5){16}}
\multiput(140,140)(-16,-26.667){4}{\line(-3,-5){12}}
\multiput(140,0)(60,100){2}{\circle*{8}}
\multiput(220,40)(60,100){2}{\circle*{8}}
\multiput(60,100)(60,100){2}{\circle*{8}}
\multiput(0,140)(60,100){2}{\circle*{8}}
\multiput(140,280)(80,-100){2}{\circle*{8}}
\multiput(80,180)(80,-100){2}{\circle*{8}}
\multiput(80,40)(120,200){2}{\circle*{8}}
\put(140,140){\circle*{12}}
\put(255,25){\vector(-1,1){40}}\put(255,240){\vector(-1,-1){42.5}}
\put(30,20){\vector(1,1){51}}\put(30,253){\line(0,-1){50}}
\multiput(30,198)(0,-10){2}{\line(0,-1){5}}\put(30,178){\vector(0,-1){15}}
\put(250,5){ ${W'''}$}
\put(260,245){${W'}$}
\put(0,0){${ W}$}
\put(0,255){${W''}$}
\put(45,115){$a_4$}\put(-23,150){$c_3$}\put(65,20){$b_3$}\put(135,-20){$c_2$}
\put(220,20){$a_1$}\put(150,60){$c_5$}\put(130,153){$z$}\put(290,133){$c_6$}
\put(202,250){$a_3$}\put(130,295){$c_4$}\put(55,255){$b_1$}\put(65,195){$a_2$}
\put(220,155){$b_4$}\put(205,83){$b_2$}\put(115,215){$c_1$}
\end{picture}}
\put(450,100){\begin{picture}(300,300)
\thicklines
\multiput(141,0)(60,100){2}{\line(2,1){80}}
\multiput(140,0)(60,100){2}{\line(2,1){80}}
\multiput(139,0)(60,100){2}{\line(2,1){80}}
\multiput(140,1)(80,40){2}{\line(3,5){60}}
\multiput(140,0)(80,40){2}{\line(3,5){60}}
\multiput(140,-1)(80,40){2}{\line(3,5){60}}
\multiput(0,141)(60,-40){2}{\line(3,5){60}}
\multiput(0,140)(60,-40){2}{\line(3,5){60}}
\multiput(0,139)(60,-40){2}{\line(3,5){60}}
\multiput(1,140)(60,100){2}{\line(3,-2){60}}
\multiput(0,140)(60,100){2}{\line(3,-2){60}}
\multiput(-1,140)(60,100){2}{\line(3,-2){60}}
\multiput(0,141)(60,-40){2}{\line(4,-5){80}}
\multiput(0,140)(60,-40){2}{\line(4,-5){80}}
\multiput(0,139)(60,-40){2}{\line(4,-5){80}}
\multiput(200,101)(80,40){2}{\line(-4,5){80}}
\multiput(200,100)(80,40){2}{\line(-4,5){80}}
\multiput(200, 99)(80,40){2}{\line(-4,5){80}}
\multiput(61,240)(60,-40){2}{\line(2,1){80}}
\multiput(60,240)(60,-40){2}{\line(2,1){80}}
\multiput(59,240)(60,-40){2}{\line(2,1){80}}
\multiput(141,0)(60,240){2}{\line(-3,2){60}}
\multiput(140,0)(60,240){2}{\line(-3,2){60}}
\multiput(139,0)(60,240){2}{\line(-3,2){60}}
\multiput(160,80)(30,-20){2}{\line(3,-2){20}}
\multiput(220,180)(30,-20){2}{\line(3,-2){20}}
\multiput(80,40)(30,15){3}{\line(2,1){20}}
\multiput(0,140)(30,15){3}{\line(2,1){20}}
\multiput(160,80)(-20,25){4}{\line(-4,5){16}}
\multiput(220,180)(-20,25){4}{\line(-4,5){16}}
\multiput(220,180)(-16,-26.667){4}{\line(-3,-5){12}}
\multiput(140,280)(-16,-26.667){4}{\line(-3,-5){12}}
\multiput(140,140)(-30,20){2}{\line(-3,2){20}}
\multiput(140,140)(-30,-15){3}{\line(-2,-1){20}}
\multiput(140,140)(20,-25){4}{\line(4,-5){16}}
\multiput(140,140)(16,26.667){4}{\line(3,5){12}}
\multiput(140,0)(60,100){2}{\circle*{8}}
\multiput(220,40)(60,100){2}{\circle*{8}}
\multiput(60,100)(60,100){2}{\circle*{8}}
\multiput(0,140)(60,100){2}{\circle*{8}}
\multiput(140,280)(80,-100){2}{\circle*{8}}
\multiput(80,180)(80,-100){2}{\circle*{8}}
\multiput(80,40)(120,200){2}{\circle*{8}}\put(140,140){\circle*{12}}
\put(270,0){\sc${W''}$}
\put(-10,-20){\sc${W'}$}
\put(260,265){\sc${W}$}
\put(20,275){\sc${W'''}$}
\put(260,20){\vector(-1,1){50}}\put(250,260){\line(-1,-1){42}}
\put(203,213){\vector(-1,-1){20}}
\put(16,6){\vector(1,1){63.5}}\put(38,260){\vector(0,-1){80}}
\put(45,115){$a_4$}\put(-23,150){$c_3$}\put(65,20){$b_3$}\put(135,-20){$c_2$}
\put(220,20){$a_1$}\put(150,60){$c_5$}\put(130,153){$z$}\put(290,133){$c_6$}
\put(202,250){$a_3$}\put(130,295){$c_4$}\put(55,255){$b_1$}\put(65,195){$a_2$}
\put(220,155){$b_4$}\put(205,83){$b_2$}\put(115,215){$c_1$}
\end{picture}}
\put(375,210){\huge\bf $=$} 
\end{picture}
\caption{ Graphical representation for the tetrahedron
  equations for interaction-round-a-cube models.}\label{TEIRC}
\end{figure}
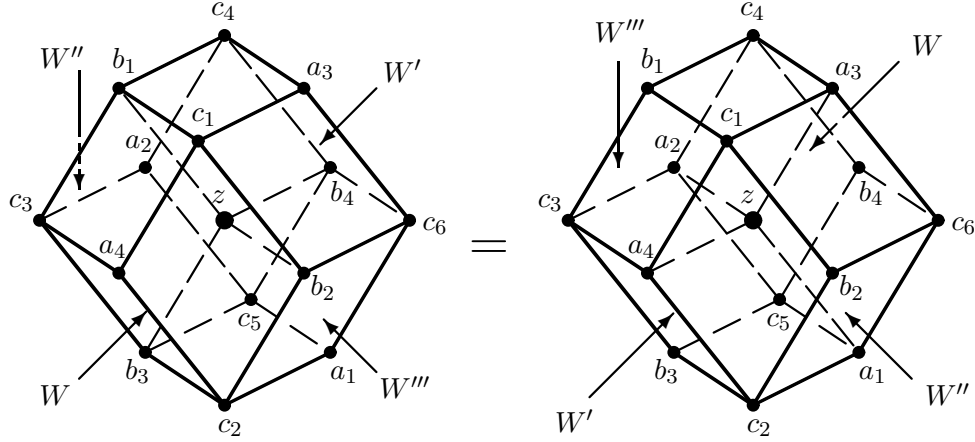

The weight function (\ref{Wweight}) satisfies the tetrahedron equation of the
form \cite{Baxter:1986phd} (see Fig.~\ref{TEIRC}),
\be
\begin{array}{l}
\ds \int_{\mathbb{R}} dz\, W(a_4|c_2,c_1,c_3|b_1,b_3,b_2|z)\,
W'(c_1|b_2,a_3,b_1|c_4,z,c_6|b_4)\\[.3cm]
\qquad\times W''(b_1|z,c_4,c_3|a_2,b_3,b_4|c_5)\,
W'''(z|b_2,b_4,b_3|c_5,c_2,c_6|a_1)=\\[.4cm]
\phantom{XXXXXXXXX}
\ds= \int_{\mathbb{R}} dz \,W'''(b_1|c_1,c_4,c_3|a_2,a_4,a_3|z)\,
W''(c_1|b_2,a_3,a_4|z,c_2,c_6|a_1)\\[.3cm]
\phantom{XXXXXXXXXX}
\qquad
\times W'(a_4|c_2,z,c_3|a_2,b_3,a_1|c_5)\,
W(z|a_1,a_3,a_2|c_4,c_5,c_6|b_4),\label{tetr}
\end{array}
\ee
where the four sets of the spectral parameters are constrained as
\begin{equation}
\T_2'=\T_2,\quad \T_2''=-\T_3,\quad \T_3''=\T_3',
\quad \T_1'''=\T_1'',\quad \T_2'''=\T_1,
\quad \T_3'''=-\T_1'\label{Tshki}
\end{equation}
and
\begin{equation}
\a_1''=\a_1'-\a_1,\quad \a_1'''=\a_3'-\a_3,
\quad \a_2'''=\a_2''-\a_2',\quad \a_3'''=\a_3''-\a_2.\label{ashki}
\end{equation}

The partition function is defined as
\be\label{part-m2}
Z'_{\mathsf M}=\int \cdots \int \prod_{\mathrm{cubes}} W(a|efg|bcd|h) \
\prod_{\mathrm{vertices}} \ da
\ee
where the first product is taken over all elementary cubes and the
integral is taken over all vertex spins of the
lattice $a, b, c, \ldots $. This model can be thought of as a
non-compact analog of the generalized Zamolodchikov model
\cite{Bazhanov:1992jq}, corresponding to an analytic continuation of
the latter to complex values of the number of spin states
$N=-b^{2}$.

A precise connection between the two partition functions \eqref{pf-m}
and  \eqref{part-m2} requires a detailed discussion of the boundary
conditions in both models which is postponed to \cite{BMS08}.
Here we only note that expression  \eqref{pf-m} is particular case of
\eqref{part-m2} corresponding to a special choice of the
field parameters $\a_1,\a_2,\a_3$ in \eqref{Wweight}.

\section{Conclusion}
In this paper we have exposed various connections between
discrete differential geometry, statistical mechanics
and quantum field theory, displaying geometric origins of algebraic structures
underlying integrability of quantum systems.

We have shown that the 3D circular lattices are associated with an
integrable discrete Hamiltonian system and constructed two different
quantizations of this system.  The resulting 3D integrable models
can be thought of as describing quantum fluctuations of the lattice
geometry. The classical geometry of the 3D circular lattices arises as
a stationary configuration giving the leading contribution to the
partition function of the quantum system in the quasi-classical
limit.

We have also obtained two solutions of the tetrahedron equation,
which naturally arise in our approach to the quantization of the circular
lattices. One of these solutions is new.
It has continuous spin variables taking values on the real line.
 This solution is connected with the modular double  of the
 $q$-oscillator algebra and can be considered as a non-compact
 counterpart of the generalized Zamolodchikov model \cite{Bazhanov:1992jq}.
The other solution of the tetrahedron equation which arose here in the
context of the circular lattices, was previously
constructed in \cite{BS05} by a purely algebraic approach.
This solution possesses
a remarkable property:
it reproduces all two-dimensional solvable models related to
finite-dimensional highest weight representations for all
quantized affine algebras $U_q(\widehat{sl}_n)$,
$n=2,3,\ldots,\infty$ (the rank $n$ coincides with the size
of one dimension of the 3D lattice).
Plausibly, a similar 3D interpretation, originating from other
simple geometrical models, also exists for the
trigonometric solutions of the Yang-Baxter equation, related with all
other infinite series of quantized affine algebras
\cite{Bazhanov:1984gu,Jimbo86} and super-algebras
\cite{Bazhanov:1987tmp}. Therefore, it might very well
be that not only the phenomenon of quantum
integrability but the quantized algebras
themselves are deeply connected with geometry.

Here we only stated our main results leaving detailed derivations
to future publications \cite{BMS08}.
There are many other questions we would also like to address there, in
particular the geometric meaning of the Poisson algebra
\eqref{poisson} and connections of the 3D circular lattices with the 2D
circle patterns \cite{BSp} on the plane or the sphere.
It would be interesting to understand underlying reasons of a
``persistent'' appearance of the $q$-oscillator algebra \eqref{q-osc1}
as a primary algebraic structure in many other important aspects
of the theory of integrable systems, such as, for example, the
construction of Baxter's ${\bf Q}$-operators \cite{BLZ97a} and
the calculation of correlation functions of the XXZ model \cite{BJMS06}.
It would also be
interesting to explore connections of our results
with the invariants of the 3D manifolds \cite{W89,RT91,TV92},
the link invariants \cite{Kash95,Mur2001,Hikami:2006},
quantization of the Techmueller space \cite{Kash98,Fok99} and
the representation theory of $U_q(sl(2|{\mathbb R})$ \cite{PT01}.
So, there are many interesting questions about the quantum integrability
still remain unanswered, but one thing is getting
is more and more clear: it is not just connected with geometry, it is
geometry itself! (though the Quantum Geometry).

\section*{Acknowledgments}
The authors thank R.J.Baxter, M.T.Batchelor, A.Doliwa, M.Jimbo, R.M.Kashaev,
T.Miwa and F.A.Smirnov for interesting discussions and remarks.
One of us (VB) thanks M.Staudacher for his hospitality at the
Albert Einstein
Institute for Gravitational Physics in Golm, where some parts of
this work have been done.
Special thanks to A.I.Bobenko for numerous important comments and 
to D. Whitehouse at the ANU Supercomputer Facility
for the professional graphics of the Miquel circles (Fig.~7).

\def\cprime{$'$} \def\cprime{$'$}


\begin{thebibliography}{10}

\bibitem{Yang:1967}
Yang, C.~N.
\newblock Some exact results for the many-body problem in one dimension with
  repulsive delta-function interaction.
\newblock Phys. Rev. Lett. {\bf 19} (1967) 1312--1315.

\bibitem{Bax72}
Baxter, R.~J.
\newblock Partition function of the eight-vertex lattice model.
\newblock Ann. Physics {\bf 70} (1972) 193--228.

\bibitem{Dri87}
Drinfel'd, V.~G.
\newblock Quantum groups.
\newblock In {\em Proceedings of the International Congress of Mathematicians,
  Vol. 1, 2 (Berkeley, Calif., 1986)}, pages 798--820, Providence, RI, 1987.
  Amer. Math. Soc.

\bibitem{Jim85}
Jimbo, M.
\newblock A $q$-difference analogue of ${U}({G})$ and the {Y}ang-{B}axter
  equation.
\newblock Lett. Math. Phys. {\bf 10} (1985) 63--69.

\bibitem{BPZ84}
Belavin, A.~A., Polyakov, A.~M., and Zamolodchikov, A.~B.
\newblock Infinite conformal symmetry in two-dimensional quantum field theory.
\newblock Nuclear Phys. B {\bf 241} (1984) 333--380.

\bibitem{BMS07a}
Bazhanov, V.~V., Mangazeev, V.~V., and Sergeev, S.~M.
\newblock Faddeev-Volkov solution of the Yang-Baxter Equation and Discrete
  Conformal Symmetry.
\newblock Nucl. Phys. {\bf B784} (2007) 234--258.

\bibitem{BMS07b}
Bazhanov, V.~V., Mangazeev, V.~V., and Sergeev, S.~M.
\newblock Exact solution of the Faddeev-Volkov model.
\newblock Phys. Lett. A {\bf 372} (2008) 1547--1550.
\newblock arXiv.org:0706.3077.

\bibitem{Volkov:1992}
Volkov, A.~Y.
\newblock Quantum {V}olterra model.
\newblock Phys. Lett. A {\bf 167} (1992) 345--355.

\bibitem{FV:1993}
Faddeev, L. and Volkov, A.~Y.
\newblock Abelian current algebra and the Virasoro algebra on the lattice.
\newblock Phys. Lett. B {\bf 315} (1993) 311--318.

\bibitem{Faddeev:1994}
Faddeev, L.
\newblock Currentlike variables in massive and massless integrable models.
\newblock In {\em Quantum groups and their applications in physics (Varenna,
  1994)}, volume 127 of {\em Proc. Internat. School Phys. Enrico Fermi}, pages
  117--135, Amsterdam, 1996. IOS.

\bibitem{BSp}
Bobenko, A.~I. and Springborn, B.~A.
\newblock Variational principles for circle patterns and Koebe's theorem.
\newblock Trans. Amer. Math. Soc. {\bf 365} (2004) 659–689.

\bibitem{St1}
Stephenson, K.
\newblock Circle packing: a mathematical tale.
\newblock Notices Amer. Math. Soc. {\bf 50} (2003) 1376–1388.

\bibitem{Zamolodchikov:1980rus}
Zamolodchikov, A.~B.
\newblock Tetrahedra equations and integrable systems in three-dimensional
  space.
\newblock Soviet Phys. JETP {\bf 52} (1980) 325--336.

\bibitem{Zamolodchikov:1981kf}
Zamolodchikov, A.~B.
\newblock Tetrahedron equations and the relativistic {S} matrix of straight
  strings in (2+1)-dimensions.
\newblock Commun. Math. Phys. {\bf 79} (1981) 489--505.

\bibitem{Baxter:1986phd}
Baxter, R.~J.
\newblock The {Y}ang-{B}axter {E}quations and the {Z}amolodchikov {M}odel.
\newblock Physica {\bf 18D} (1986) 321--247.

\bibitem{Bazhanov:1992jq}
Bazhanov, V.~V. and Baxter, R.~J.
\newblock New solvable lattice models in three-dimensions.
\newblock J. Statist. Phys. {\bf 69} (1992) 453--585.

\bibitem{Bazhanov:1993j}
Bazhanov, V.~V. and Baxter, R.~J.
\newblock Star triangle relation for a three-dimensional model.
\newblock J. Statist. Phys. {\bf 71} (1993) 839--864.

\bibitem{Kashaev:1993ijmp}
Kashaev, R.~M., Mangazeev, V.~V., and Stroganov, Y.~G.
\newblock Spatial symmetry, local integrability and tetrahedron equations in
  the {B}axter-{B}azhanov model.
\newblock Int. J. Mod. Phys. A {\bf 8} (1993) 587.

\bibitem{Korepanov:1993jsp}
Korepanov, I.~G.
\newblock Tetrahedral {Z}amolodchikov algebras corresponding to {B}axter's
  ${L}$-operators.
\newblock J. Stat. Phys. {\bf 71} (1993) 85--97.

\bibitem{KashaevKorepanovSergeev}
Kashaev, R.~M., Korepanov, I.~G., and Sergeev, S.~M.
\newblock The functional tetrahedron equation.
\newblock Teoret. Mat. Fiz. {\bf 117} (1998) 370--384.

\bibitem{BS05}
Bazhanov, V.~V. and Sergeev, S.~M.
\newblock Zamolodchikov's tetrahedron equation and hidden structure of quantum
  groups.
\newblock J. Phys. A {\bf 39} (2006) 3295--3310.

\bibitem{BKMS}
Bazhanov, V.~V., Kashaev, R.~M., Mangazeev, V.~V., and Stroganov, Y.~G.
\newblock $({Z}\sb {N}\times)\sp {n-1}$ generalization of the chiral {P}otts
  model.
\newblock Comm. Math. Phys. {\bf 138} (1991) 393--408.

\bibitem{Date:1990bs}
Date, E., Jimbo, M., Miki, K., and Miwa, T.
\newblock Generalized chiral {P}otts models and minimal cyclic representations
  of ${U}_q(gl(n,{C}))$.
\newblock Commun. Math. Phys. {\bf 137} (1991) 133--148.

\bibitem{Bob96}
Bobenko, A.~I.
\newblock Discrete conformal maps and surfaces.
\newblock In {\em Symmetries and integrability of difference equations
  (Canterbury, 1996)}, volume 255 of {\em London Math. Soc. Lecture Note Ser.},
  pages 97--108. Cambridge Univ. Press, Cambridge, 1999.

\bibitem{Lame1859}
Lam\'e, G.
\newblock {\em Lecons sur la The\'eorie des coorden\'ees curvilignes et leurs
  diverses applications.}
\newblock Mallet-Bachalier, Paris, 1859.

\bibitem{Darboux}
Darboux, G.
\newblock {\em Le\c{c}ons sur les syst\'emes orthogonaux et les coordonn\'ees
  curvilignes}, volume I-IV.
\newblock Gauthier-Villars, Paris, 1910.

\bibitem{ZakharovManakov}
Zakharov, V.~E. and Manakov, S.~V.
\newblock Construction of multidimensional nonlinear integrable systems and
  their solutions.
\newblock Funct. Anal. Appl. {\bf 19} (1985) 89--101.

\bibitem{Kri96}
Krichever, I.~M.
\newblock Algebraic-geometric {$n$}-orthogonal curvilinear coordinate systems
  and the solution of associativity equations.
\newblock Funct. Anal. Appl. {\bf 31} (1997) 25--39.

\bibitem{BobenkoPinkall}
Bobenko, A. and Pinkall, U.
\newblock Discrete isothermic surfaces.
\newblock J. Reine Angew. Math. {\bf 475} (1996) 187--208.

\bibitem{DoliwaSantini}
Doliwa, A. and Santini, P.~M.
\newblock Multidimensional quadrilateral lattices are integrable.
\newblock Phys. Lett. A {\bf 233} (1997) 365--372.

\bibitem{AdlerBobenkoSuris}
Adler, V.~E., Bobenko, A.~I., and Suris, Y.~B.
\newblock Classification of integrable equations on quad-graphs. {T}he
  consistency approach.
\newblock Comm. Math. Phys. {\bf 233} (2003) 513--543.

\bibitem{KonopelchenkoSchief}
Konopelchenko, B.~G. and Schief, W.~K.
\newblock Menelaus' theorem, {C}lifford configurations and inversive geometry
  of the {S}chwarzian {KP} hierarchy.
\newblock J. Phys. A {\bf 35} (2002) 6125--6144.

\bibitem{CDS97}
Cieslinski, J., Doliwa, A., and Santini, P.~M.
\newblock The Integrable Discrete Analogues of Orthogonal Coordinate Systems
  are Multidimensional Circular Lattices.
\newblock Phys. Lett. {\bf A235} (1997) 480--488.

\bibitem{KS98}
Konopelchenko, B.~G. and Schief, W.~K.
\newblock Three-dimensional integrable lattices in {E}uclidean spaces:
  conjugacy and orthogonality.
\newblock R. Soc. Lond. Proc. Ser. A Math. Phys. Eng. Sci. {\bf 454} (1998)
  3075--3104.

\bibitem{DMS98}
Doliwa, A., Manakov, S., and Santini, P.
\newblock $\overline{\partial}$-reductions of the multidimensional
  quadrilateral lattice: the multidimensional circular lattice.
\newblock Commun. Math. Phys. {\bf 196} (1998) 1--18.

\bibitem{BoSurUMN07}
Bobenko, A.~I. and Suris, Y.~B.
\newblock On discretization principles for differential geometry. {T}he
  geometry of spheres.
\newblock Uspekhi Mat. Nauk {\bf 62} (2007) 3--50.

\bibitem{BoSur05}
Bobenko, A.~I. and Suris, Y.~B.
\newblock Discrete differential geometry. Consistency as integrability, 2005.
\newblock Preliminary version of a book. arXiv:math/0504358.

\bibitem{BMS08}
Bazhanov, V.~V., Mangazeev, V.~V., and Sergeev, S.~M.
\newblock Integrable Systems and Quantum Discrete Geometry.
\newblock In preparation., 2008.

\bibitem{Kas96}
Kashaev, R.
\newblock {On discrete three-dimensional equations associated with the local
  Yang-Baxter relation}.
\newblock Letters in Mathematical Physics {\bf 38} (1996) 389--397.

\bibitem{BogdanovKonopelchenko}
Bogdanov, L.~V. and Konopelchenko, B.~G.
\newblock Lattice and {$q$}-difference {D}arboux-{Z}akharov-{M}anakov systems
  via {$\overline\partial$}-dressing method.
\newblock J. Phys. A {\bf 28} (1995) L173--L178.

\bibitem{DS00}
Doliwa, A. and Santini, P.~M.
\newblock The symmetric, $d$-invariant and Egorov reductions of the
  quadrilateral lattice.
\newblock J. Geom. Phys. {\bf 36} (2000) 60--102.

\bibitem{Miquel}
Miquel, A.
\newblock Th\'eor\`emes sur les intersections des cercles et des sph\`eres.
\newblock J. Math. Pur. Appl. (Liouville J.) {\bf 3} (1838) 517--522.

\bibitem{KMS1997}
Korepanov, I.~G., Maillard, J.-M., and Sergeev, S.~M.
\newblock Classical limit for a {$3$}{D} lattice spin model.
\newblock Phys. Lett. A {\bf 232} (1997) 211--216.

\bibitem{Schmud94}
Schm{\"u}dgen, K.
\newblock Integrable operator representations of {${\bf R}\sp 2\sb q$}, {$X\sb
  {q,\gamma}$} and {${\rm SL}\sb q(2,{\bf R})$}.
\newblock Comm. Math. Phys. {\bf 159} (1994) 217--237.

\bibitem{Faddeev:1995}
Faddeev, L.~D.
\newblock Discrete Heisenberg-Weyl group and modular group.
\newblock Lett. Math. Phys. {\bf 34} (1995) 249--254.

\bibitem{Faddeev:1999}
Faddeev, L.
\newblock Modular double of a quantum group.
\newblock In {\em Conf\'{e}rence Mosh\'{e} Flato 1999, Vol. I (Dijon)},
  volume~21 of {\em Math. Phys. Stud}, pages 149--156, Dordrecht, 2000. Kluwer
  Acad. Publ.
\newblock math/9912078.

\bibitem{PT99}
Ponsot, B. and Teschner, J.
\newblock Liouville bootstrap via harmonic analysis on a noncompact quantum
  group.
\newblock (1999).
\newblock hep-th/9911110.

\bibitem{milnor}
Milnor, J.
\newblock Hyperbolic geometry: the first 150 years.
\newblock Bull. Amer. Math. Soc. (N.S.) {\bf 6} (1982) 9--24.

\bibitem{Bazhanov:1984gu}
Bazhanov, V.~V.
\newblock Trigonometric solution of triangle equations and classical {L}ie
  algebras.
\newblock Phys. Lett. {\bf B159} (1985) 321--324.

\bibitem{Jimbo86}
Jimbo, M.
\newblock Quantum {$R$} matrix for the generalized {T}oda system.
\newblock Comm. Math. Phys. {\bf 102} (1986) 537--547.

\bibitem{Bazhanov:1987tmp}
Bazhanov, V.~V. and Shadrikov, A.~G.
\newblock Quantum triangle equations and simple {L}ie-superalgebras.
\newblock Theor. Math. Phys. {\bf 73} (1987) 1303--1312.

\bibitem{BLZ97a}
Bazhanov, V.~V., Lukyanov, S.~L., and Zamolodchikov, A.~B.
\newblock Integrable structure of conformal field theory. {I}{I}. ${
  Q}$-operator and {D}{D}{V} equation.
\newblock Comm. Math. Phys. {\bf 190} (1997) 247--278.
\newblock [{\tt hep-th/9604044}].

\bibitem{BJMS06}
Boos, H., Jimbo, M., Miwa, T., Smirnov, F., and Takeyama, Y.
\newblock Hidden {G}rassmann structure in the {$XXZ$} model.
\newblock Comm. Math. Phys. {\bf 272} (2007) 263--281.
\newblock hep-th/0606280.

\bibitem{W89}
Witten, E.
\newblock Quantum field theory and the {J}ones polynomial.
\newblock Comm. Math. Phys. {\bf 121} (1989) 351--399.

\bibitem{RT91}
Reshetikhin, N. and Turaev, V.~G.
\newblock Invariants of {$3$}-manifolds via link polynomials and quantum
  groups.
\newblock Invent. Math. {\bf 103} (1991) 547--597.

\bibitem{TV92}
Turaev, V.~G. and Viro, O.~Y.
\newblock State sum invariants of {$3$}-manifolds and quantum {$6j$}-symbols.
\newblock Topology {\bf 31} (1992) 865--902.

\bibitem{Kash95}
Kashaev, R.~M.
\newblock A link invariant from quantum dilogarithm.
\newblock Modern Phys. Lett. A {\bf 10} (1995) 1409--1418.

\bibitem{Mur2001}
Murakami, H. and Murakami, J.
\newblock The colored Jones polynomials and the simplicial volume of a knot.
\newblock Acta Mathematica {\bf 186} (2001) 85--104.

\bibitem{Hikami:2006}
Hikami, K.
\newblock Generalized Volume Conjecture and the A-Polynomials -- the
  Neumann-Zagier Potential Function as a Classical Limit of Quantum Invariant.
\newblock J. Geom. Phys. {\bf 57} (2007) 1895--1940.
\newblock arXiv.org:math.QA/0604094.

\bibitem{Kash98}
Kashaev, R.~M.
\newblock Quantization of {T}eichm\"uller spaces and the quantum dilogarithm.
\newblock Lett. Math. Phys. {\bf 43} (1998) 105--115.

\bibitem{Fok99}
Fok, V.~V. and Chekhov, L.~O.
\newblock Quantum {T}eichm\"uller spaces.
\newblock Teoret. Mat. Fiz. {\bf 120} (1999) 511--528.

\bibitem{PT01}
Ponsot, B. and Teschner, J.
\newblock Clebsch-{G}ordan and {R}acah-{W}igner coefficients for a continuous
  series of representations of {$U_q({sl}(2,{\mathbb R}))$}.
\newblock Comm. Math. Phys. {\bf 224} (2001) 613--655.

\end{thebibliography}

\end{document}